\documentclass[lettersize,journal]{IEEEtran}
\usepackage{amsmath,amsfonts}
\usepackage{algorithmic}
\usepackage{array}
\usepackage[caption=false,font=normalsize,labelfont=sf,textfont=sf]{subfig}
\usepackage{textcomp}
\usepackage{stfloats}
\usepackage{url}
\usepackage{verbatim}
\usepackage{graphicx}
\usepackage{svg}
\usepackage{booktabs}
\usepackage{multirow}
\usepackage{makecell}
\usepackage{svg}
\usepackage{cite}
\def\BibTeX{{\rm B\kern-.05em{\sc i\kern-.025em b}\kern-.08em
    T\kern-.1667em\lower.7ex\hbox{E}\kern-.125emX}}
\usepackage{balance}
\begin{document}
\title{Digital Twin Empowered In-Vehicular Channel Modeling \\ and Wireless Planning in the Terahertz Band}
\author{Mingjie Zhu, Yejian Lyu,~\IEEEmembership{Member,~IEEE} and Chong Han~\IEEEmembership{Senior~Member,~IEEE}
\thanks{
Mingjie~Zhu and Yejian~Lyu are with the Terahertz Wireless Communications (TWC) Laboratory, Shanghai Jiao Tong University, Shanghai 200240, China (email:~\{mingjie.zhu, yejian.lyu\}@sjtu.edu.cn). 

Chong~Han is with the Terahertz Wireless Communications (TWC) Laboratory, and Cooperative Medianet Innovation Center (CMIC), School of Information Science and Electronic Engineering, Shanghai Jiao Tong University, Shanghai 200240, China (email:~chong.han@sjtu.edu.cn). 
}
}

\maketitle
\begin{abstract}
Vehicle-to-everything (V2X) technology has emerged as a key enabler of intelligent transportation systems, while the Terahertz (THz) band offers abundant spectrum resources to support ultra-high-speed and low-latency V2X communications. This paper investigates the in-vehicle wireless channel in the 300~GHz band. First, channel measurement based on vector-network-analyzer (VNA) is conducted under typical V2X scenarios, including with/without human, and window-on/off cases. Then, a digital twin (DT) of the vehicle is constructed from high-resolution point cloud data and a measurement-based material property database. The DT is integrated into an open-source ray-tracing (RT) simulator, Sionna, to model multipath propagation. The DT-empowered simulation results are analyzed and validated with the measurement data, showing strong agreement and validating the feasibility. Finally, a hybrid ray-tracing-statistic channel model is established, combining the RT results and measurement data. Leveraging the validated model, further wireless planning is carried out, including signal-to-interference-plus-noise ratio (SINR) analysis, coverage probability evaluation, and optimal transmitter (Tx) placement. These findings provide valuable insights for the design and deployment of future THz in-vehicle communication systems.
\end{abstract}


\section{Introduction}
\label{section:intro}
\IEEEPARstart{W}{ith} the rapid growth of intelligent vehicles, the communication between vehicles and anything entails mounting significance. More recently, the growing popularity of electric vehicles equipped with large screens has driven a strong demand for ultra-fast vehicle-to-everything (V2X)~\cite{9345798, 7982949} communication to support bandwidth-intensive applications. These applications demand high-capacity, low-latency, and highly reliable wireless channels to ensure seamless data exchange and real-time responsiveness. To fulfill this demand, the Terahertz (THz) band, with its abundant spectrum resources, is considered a promising technology enabler, offering enhanced reliability and ultra-low latency across diverse environments~\cite{AKYILDIZ201416,9794668,9887921,6005340}. 
In particular, THz-based in-vehicle communication can support next-generation intelligent transportation systems, complementing or even surpassing current technologies such as Bluetooth or Wi-Fi for high-speed intra-vehicle connectivity.
The foundation of V2X system design lies in a precise understanding of the underlying V2X wireless channel, which can be obtained from digital twin~(DT). A DT is a digital replica of a physical object or environment that preserves intrinsic geometric and material properties, enabling precise input for RT simulations~\cite{10188847, 9120192}. RT is a technique of depicting a channel by tracing the paths of electromagnetic waves as they reflect, scatter, or diffract from the Tx to the Rx~\cite{9277601}. With detailed environmental information provided by DT, it is effective for RT simulators to determine reflection, diffusion, scattering, and estimate losses.


Accordingly, extensive research has been conducted to model and characterize V2X channels in the microwave and millimeter-wave bands~\cite{10938769,9316947, 8473696,9403881,9930640}. The impact of the random walk process of the clusters and the velocity variations of the communication terminals on these statistical properties is studied in ~\cite{9316947}.  In~\cite{8473696}, obstructed V2V channels at $5$~GHz were studied using measurement-calibrated ray-tracing (RT) simulations across various terrains and antenna setups, emphasizing the impact of obstructing vehicles.
Till date, limited channel measurement and modeling efforts were reported at sub-THz frequencies. A measurement campaign at $300$~GHz was conducted in typical V2V scenarios~\cite{9403881}, providing foundational insights for modeling future low-THz vehicular channels. Furthermore, V2I communications were explored in~\cite{9930640} using both RT simulations and real-world measurements, reinforcing the feasibility of THz-band V2X systems. 

Still as an open research problem, the in-vehicle scenario is not mentioned in the above works, and the high-precision DT of vehicles is not utilized.
Deterministic simulation aided by DT is the other aspect of V2X communication, often implemented by RT.
Although various RT simulators have been studied~\cite{10901773}, few have addressed simulations specifically in the THz range due to the lack of material properties in THz bands. 
Some studies have conducted THz-band RT-based measurement campaigns~\cite{7982699,10327689, 10609428,10663379,9466322,10266598}. For instance, a measurement-based stochastic model is presented in~\cite{7982699}, which is calibrated with RT. ~\cite{10327689}. In~\cite{10327689}, a ray-tracing-statistical hybrid model is proposed for wireless propagation above 200~GHz. The hybrid channel model approach is also presented in a data center scenario~\cite{10609428}. Comprehensive double-directional channel measurements at 300 GHz in various usage scenarios in corridor environments in the study~\cite{10663379}, and a quasi-deterministic (QD) model is proposed. Another study in~\cite{9466322} extends the hybrid RT-statistical model in the low-THz band. A RT simulator for indoor RT simulations was employed in~\cite{10266598}, achieving strong agreement with measurements. However, research on V2X THz channel modeling remains scarce. 
To evaluate the performance of the V2X communication system and derive the best antenna deployment, system-level communication is also commonly analyzed alongside, e.g., in a wireless network\cite{10839258}.
Interference, coverage, and optimal Tx placement have all been studied in prior work~\cite{10178074}. In~\cite{9314195, 9174945}, the optimal distribution of APs in the network in UAV scenarios is focused on, but whether it is suitable for the V2X scenario is not mentioned.
Three challenges of V2X channel modeling are identified and summarized as follows. 
First, comprehensive channel measurements are required across diverse scenarios to build a complete multi-scenario V2X channel model.
Second, although the integration of RT and measurements is studied for various scenarios, a hybrid channel model under V2X scenarios has not yet been established and the application of the hybrid model is limited.
Third, there is a lack of system-level analysis methods and wireless planning strategies tailored for V2X scenarios.


In our preliminary and shorter version~\cite{globecom}, an in-vehicle channel model is proposed, a point cloud is integrated into an RT simulator to derive multipath component (MPC) parameters for validation, and interference and coverage are analyzed for the single Tx case. to acquire the optimal Tx position. 
In this work, DT of more scenarios is established to establish a hybrid in-vehicle channel model of three cases.
Furthermore, wireless planning for the in-vehicle scenario is performed, especially for the multiple transmitter (multi-Tx) case. 
The main contributions of our work can be summarized as follows.

\begin{itemize}
    \item A DT is generated to replicate the physical environment, by combining realistic 3D models and measured THz material properties. Three typical scenarios, including with/without human, and with/without window scenarios are measured to comprehensively characterize the propagation of both LoS and NLoS paths. The material database is set up through THz-TDS. The accuracy of the DT is validated by comparing RT simulation results with actual measurement data in the three typical scenarios.

    \item We develop a material-oriented hybrid channel model for the in-vehicle scenario, joining the MPCs extracted from the measured channel impulse response (CIR) through peak detection and the RT MPCs. This model includes deterministic components from ray tracing and stochastic components from diffuse scattering. The stochastic components are characterized by their cumulative distribution function (CDF). From the hybrid channel model, the material can be identified from MPCs to distinguish the material of in-vehicle reflection surfaces.

    \item DT enables efficient SINR computation for a large number of Rx locations, automatic generation of coverage maps for various antenna placements, and rapid calculation of system rate. Non-gradient-based optimization is performed empowered by DT, involving single-T and multi-Tx setups. The effects of additional Txs on interference and system performance are studied, and the optimal single-Tx placement is determined.

\end{itemize}

The remainder of the paper is organized as follows. The digital twin, together with its physical twin is illustrated in Section~\ref{section: dt}. The data processing methods are presented in Section~\ref{section:Channelmodel}. The hybrid channel model of the V2X scenario is illustrated in Section~\ref{section:validation}. The wireless planning methods of interference, coverage, and rate, along with the optimization of Tx position, are explained in Section~\ref{section:WP}. Finally, Section~\ref{section:conclusion} draws the conclusion.

\section{Digital Twin}
In this section, a DT framework is introduced to establish a connection between the physical measurement environment and its virtual counterpart. The physical part describes the real-world scenario, including the measurement setup and experimental configuration, while the digital part focuses on the construction of a digital map and the implementation of an RT-based channel simulation.
\label{section: dt}

\subsection{Physical Twin: Measurement Setups}
The physical twin represents the real-world environment where the THz channel measurements are conducted. It includes the in-vehicle measurement scenario and the channel sounding system, which provide the entitative basis for building and validating the digital model.

\label{section:md}

\subsubsection{Sounder and measurement setup}
\begin{figure}[t]
\centering
\subfloat[]{
    \includegraphics[width=0.83\columnwidth]{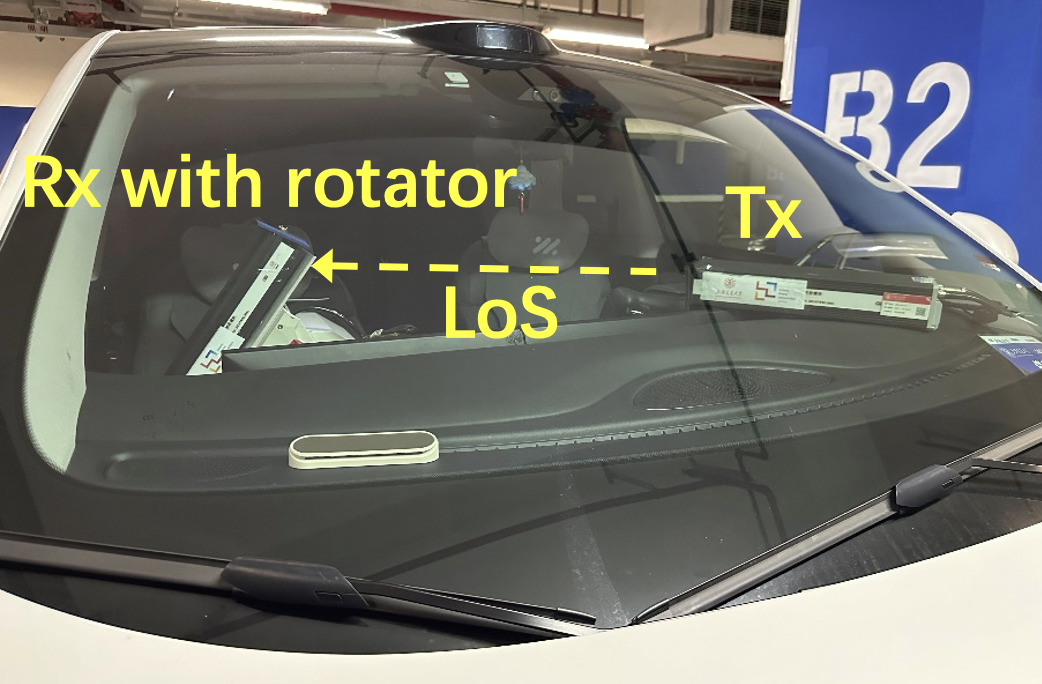}
}
\hfill
\subfloat[]{
    \includegraphics[width=0.93\columnwidth]{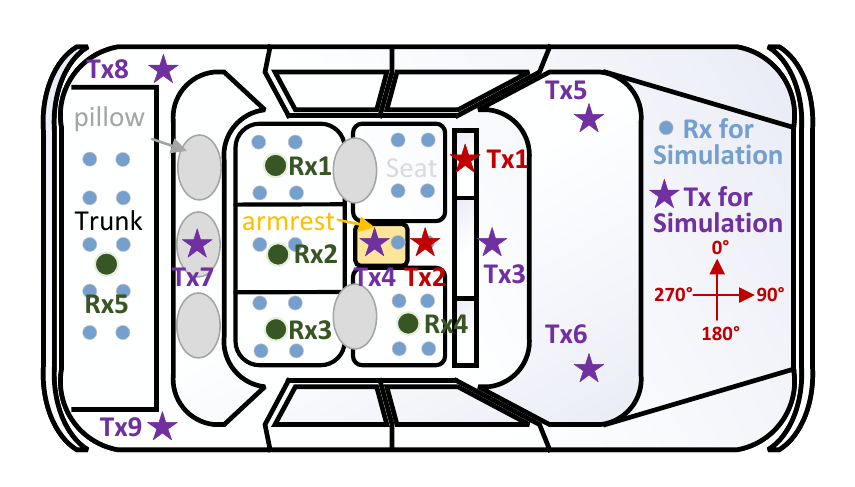}
}
\caption{Setups and scenario. (a) Real image. (b) Schematic diagram of in-vehicle scenario.}
\label{fig:scenario}
\end{figure}

\begin{table}[h]
    \caption{Measurement Configuration.}
    \label{tab:cfg}
    \centering
    \begin{tabular}{cc}
    \toprule
        Parameter & Value \\
    \midrule
        Frequancy Range & $290-310$~GHz \\
        Frequqncy Point & $2001$ \\
        IF Bandwidth & $1$~kHz \\
        Transmitted Power & $10$~dBm  \\
        Tx \& Rx Attenna Type & Horn \\
        Azimuth Rotation Step & $10^{\circ}$ \\
        In-vehicular azimuth Rotation  Range & $[0^{\circ}, 360^{\circ}]$ \\
        Zenith Rotation Step & $10^{\circ}$ \\
        In-vehicular zenith Rotation Range & $[-40^{\circ}, 60^{\circ}]$ \\
    \bottomrule
    \end{tabular}
\end{table}

Fig.~\ref{fig:scenario}~(a) shows a photograph of the measurement setup, where a vector-network-analyzer (VNA)-based channel sounder is used to capture the channel frequency responses (CFRs) inside the vehicle. The measurement is conducted in the $290$-$310$~GHz range. The broad bandwidth of $20$~GHz brings a delay resolution of $0.05$~ns, ensuring the range accuracy of $1.5$~cm. A directional scanning scheme is employed at the Rx, covering an azimuthal range of $0^{\circ}$ to $360^{\circ}$ and a zenith range of $-40^{\circ}$ to $60^{\circ}$, both with a rotation step of $10^{\circ}$, enabling the acquisition of spatial channel profiles.
Additionally, Tx is fixed and Rx is set on a rotation platform to capture the spatial channel profiles in the azimuth and zenith plane. Note that the azimuthal radius (i.e., the distance between
the phase center of the antenna to the rotation center) is 0.2 m. Before the measurements, back-to-back calibration is performed to eliminate system responses. The measurement
configurations are summarized in Table~\ref{tab:cfg}.

\subsubsection{Measurement Scenarios}
\begin{table*}[]
    \caption{Cases of Digital Twin.}
    \label{tab:cfv}
    \centering
    \begin{tabular}{>{\centering\arraybackslash}m{2.5cm}ccc}
    \toprule
       \textbf{Cases} & \textbf{CASE A} & \textbf{CASE B} & \textbf{CASE C} \\
    \midrule
        \makecell[c]{Diagram\\ \\} &
        \includegraphics[width=0.45\columnwidth]{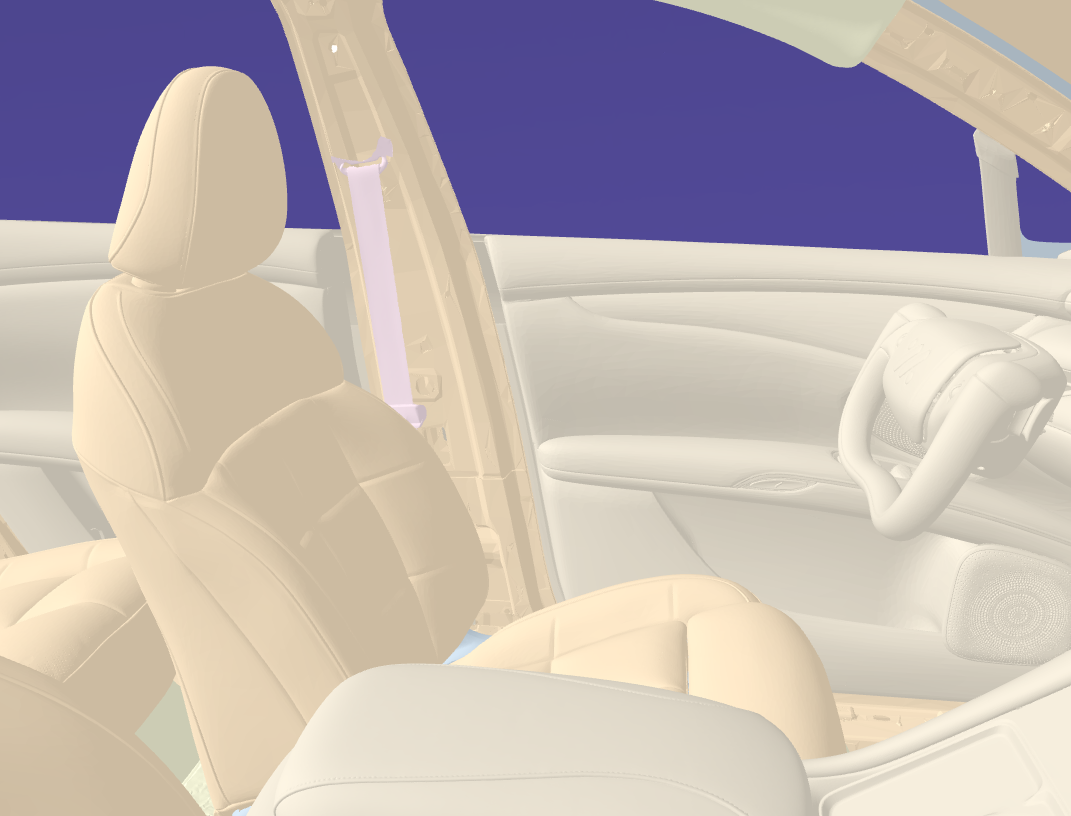} & 
        \includegraphics[width=0.45\columnwidth]{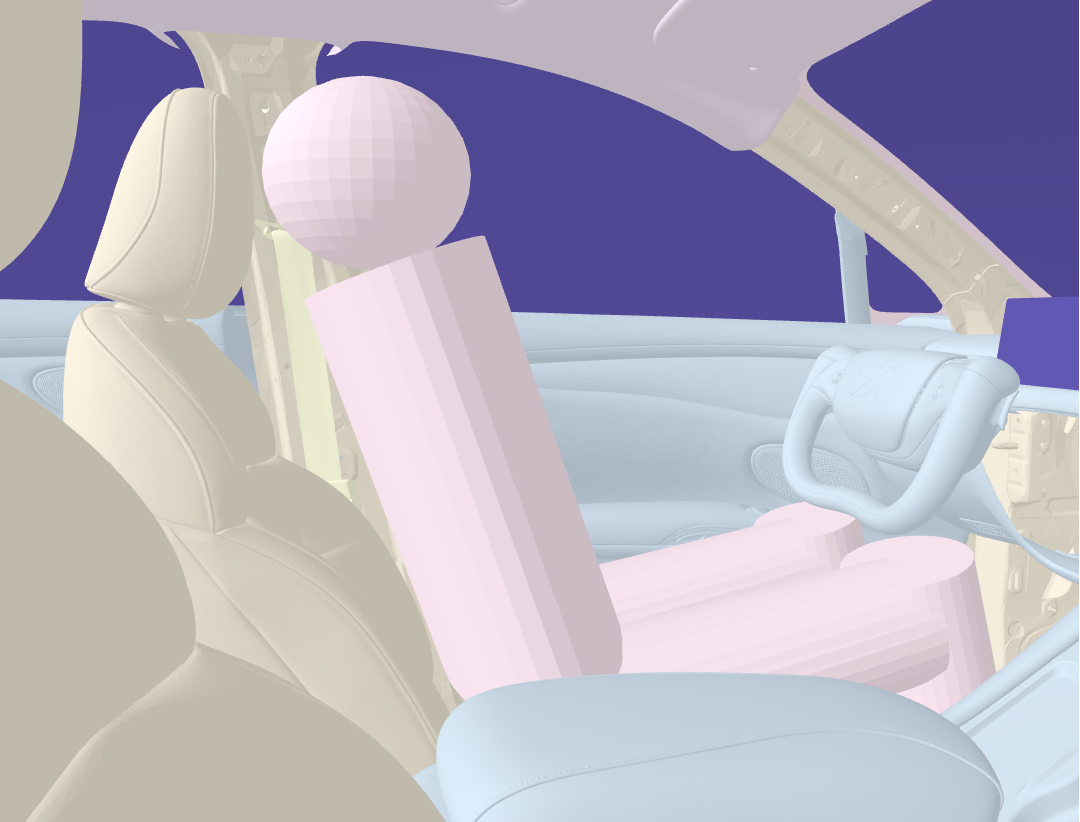} & 
        \includegraphics[width=0.45\columnwidth]{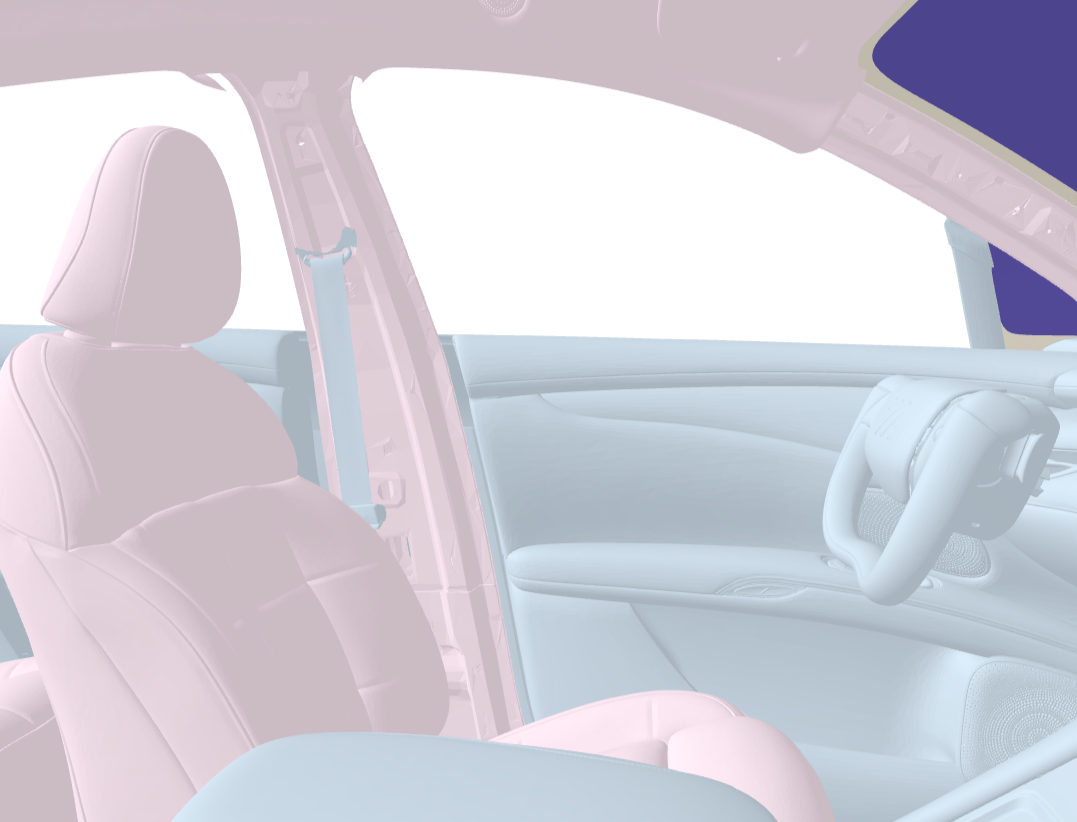} \\
        \makecell[c]{Real scene} &
        \includegraphics[width=0.45\columnwidth]{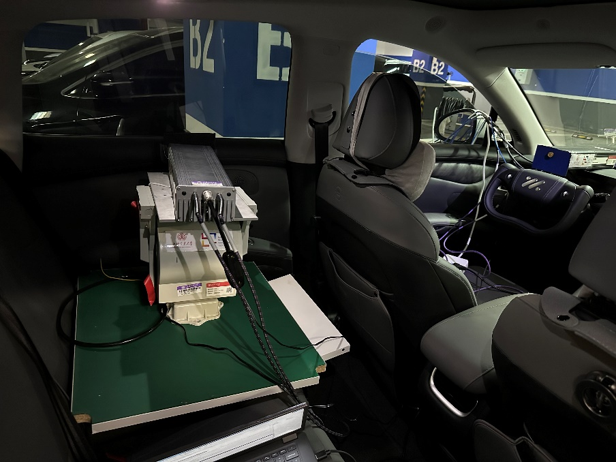} & 
        \includegraphics[width=0.45\columnwidth]{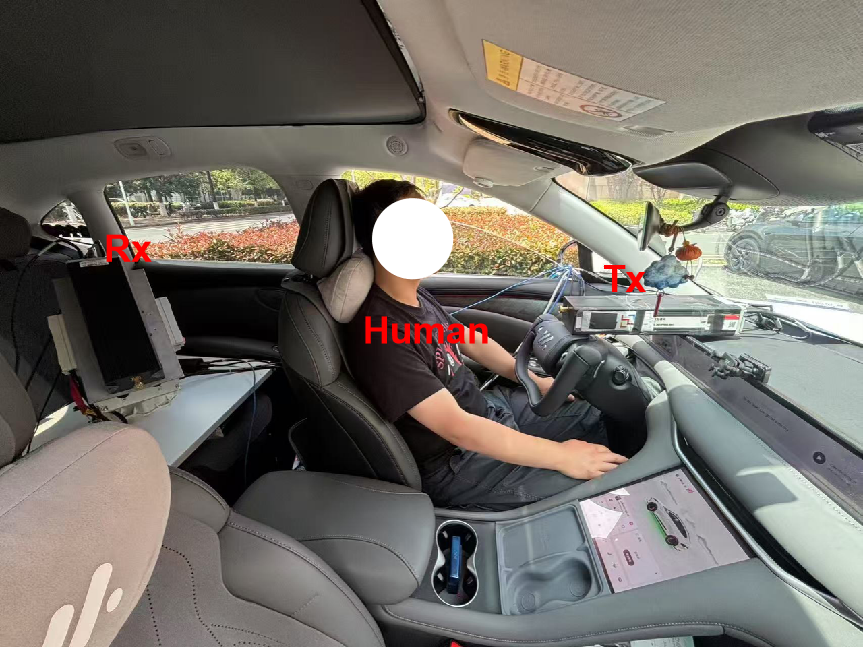} & 
        \includegraphics[width=0.45\columnwidth]{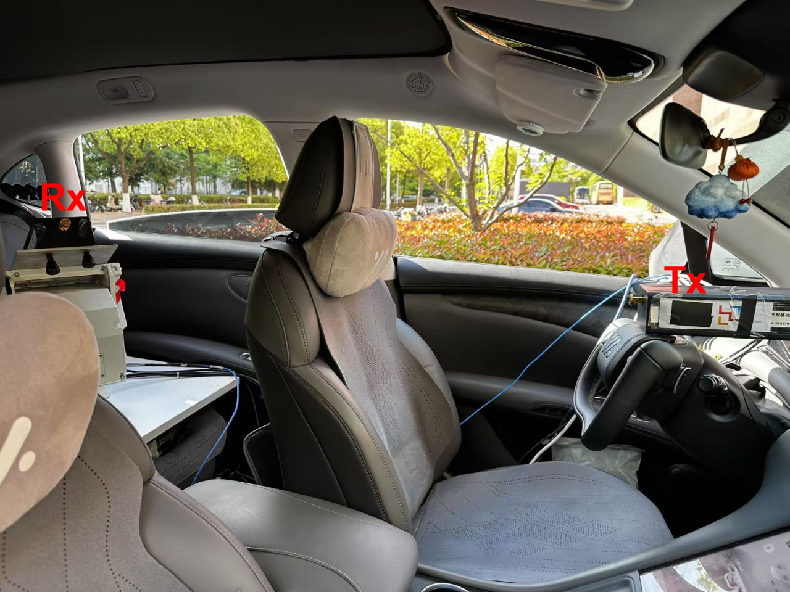} \\
        Feature & Without human/with window & With human & Without window \\
    \bottomrule
    \end{tabular}
\end{table*}

A series of V2X channel measurement experiments is conducted inside a vehicle. The tested vehicle is approximately $5$~m in length and $2$~m in width. Seats are the main blockage in the vehicle, and windows act as the main reflection surface. In the in-vehicle scenario, three sets of distinct channel measurements are carried out, which are summarized in Table~\ref{tab:cfv}, namely the with and without passengers cases, and window-on and -off cases respectively. 
Among all the cases, Tx and Rx are set at the same positions.
The top view of the vehicle, including the positions of the vehicle components as well as the Tx and Rx, is presented Fig.~\ref{fig:scenario} (b). Tx~1-4 is set right above the direction wheel, on the center armrest, above the infotainment screen, and under the ceiling. In the measurement campaign, the Tx is fixed at Tx~1 and Tx~2. The other Txs are potential Tx positions and are compared in the subsequent RT simulation. 

\subsection{Digital Twin: Simulation Framework}
\begin{figure}[t]
\centering
\includegraphics[width=1.0\columnwidth]{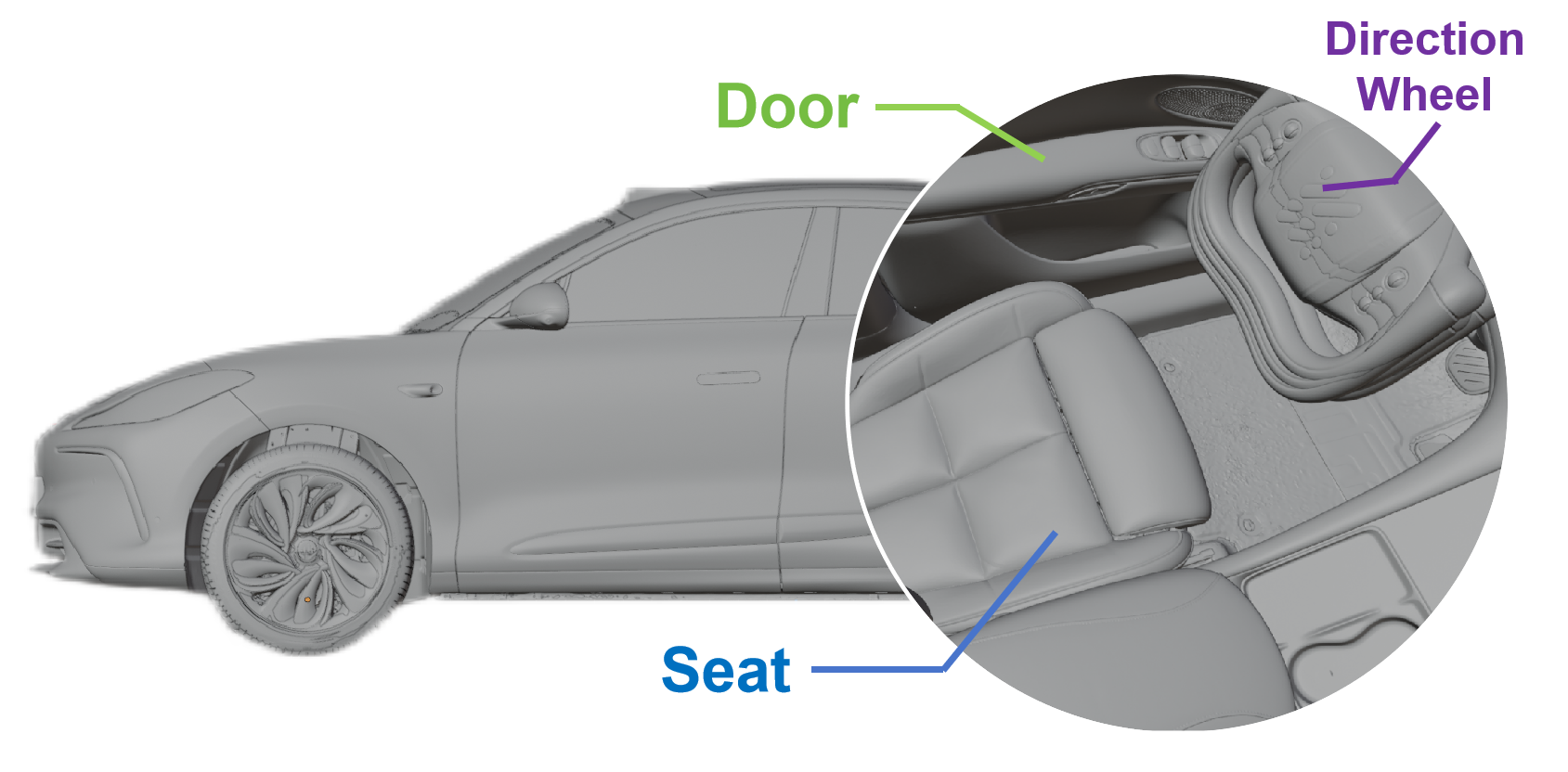}
\caption{Point cloud results of the scenario.}
\label{fig:schematic}
\end{figure}
The DT is employed to link the MPCs with environmental information, facilitating intelligent optimization and real-time control through the integration of measurement feedback and software analysis. DT results can reconstruct the vehicle scenario and handle complex circumstances, especially under tremendous repetition. 
To accomplish the DT of the vehicle scenario, a high-precision vehicle model is developed to accurately represent both the geometry and material composition of the vehicle. This process involves acquiring the structural information of the vehicle, rendering it as a point cloud, and annotating various components with corresponding material properties. The point cloud data captures the 3D spatial structure of the vehicle, and several techniques can be employed to generate such data for an object or scene of interest~\cite{10666097}. In this work, LiDAR is utilized to capture the real-world geometry of the vehicle. The resulting point cloud, as illustrated in Fig.~\ref{fig:schematic}, includes all key components such as the body-in-white (BIW), doors, seats, and windows. In addition, the material characteristics of different vehicle components are measured using THz time-domain spectroscopy (THz-TDS)~\cite{10999208}. By integrating the geometric and material data, a comprehensive DT of the in-vehicle environment is constructed. 


RT is a promising approach for deterministic channel modeling to realize DT. The open-source software Sionna~\cite{10666097,10584599} is employed for the RT simulations. Point cloud and material database of the vehicle are imported into the simulator, i.e., Sionna, enabling the computation of all path parameters between the Tx and Rx, including delay, azimuth, and zenith angle of arrival, and received power. These parameters are further processed to generate statistic diagrams for analysis.
Each of the three simulated scenarios in 
table~\ref{tab:cfv} is meticulously aligned with its physical counterpart, collectively establishing a one-to-one digital twin that faithfully mirrors the real-world vehicle environment.


In the RT simulation, path parameters are computed to derive channel coefficients, delays, and angular information. PADPs are then generated for comparison.
It is noteworthy that, due to the fact that the Rx is placed on a rotating platform during the measurement experiment, the actual receiving point changed continuously as the Rx moved. To mitigate the error introduced by this phenomenon, we no longer use a single Rx in the simulation system. Instead, multiple Rxs are deployed, each of which only filters MPCs within a specific azimuth range. 
Each Rx only accepts MPCs within $\pm 45^\circ$ of its own azimuth. By combining the MPCs collected by the four Rxs, a complete set of MPCs can be obtained. The PADP generated using this method can significantly reduce the error caused by the platform's rotation, making the comparison between simulation and measurement data more convincing.

\begin{figure}[t]
\centering
\includegraphics[width=0.9\columnwidth]{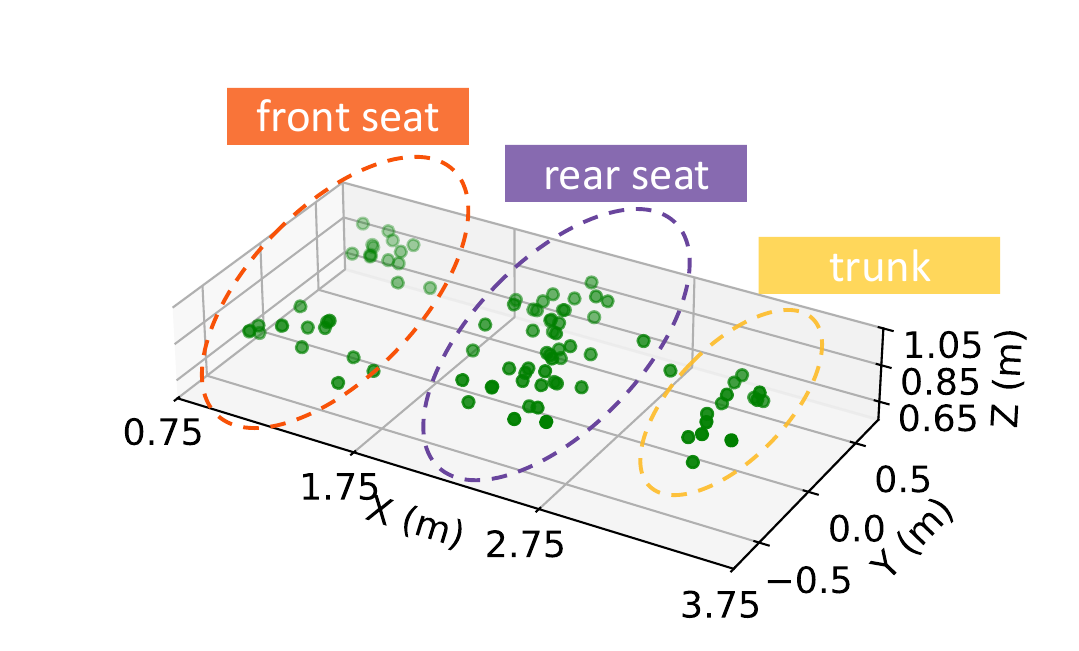}
\caption{Generated Rx Positions.}
\label{fig:RxP}
\end{figure}

The in-vehicle model can be abstracted as a physical-layer model with a fixed access point (AP) and multiple user equipment (UE). Therefore, analyzing interference and coverage in this system is necessary. Our previous work~\cite{9247469} proposed methods for calculating mean interference power and coverage probability. Similarly, these methods can be applied to the in-vehicle scenarios. 
In the wireless planning phase, additional Tx positions, i.e., Tx~2 to Tx~7, are tested following the validation experiment to assess performance differences. As shown in Fig.~\ref{fig:RxP}, a group of data points is randomly generated following a normal distribution to represent the Rx positions used for coverage and interference analysis. The $80$ Rx positions exhibit greater randomness and aim to provide coverage across as many locations within the vehicle as possible. Metrics such as interference, coverage, and data rate are analyzed across different Tx placements to identify the optimal T location.
Besides the single-T case, the circumstance of two Txs functioning in parallel is considered. Tx\,4 and Tx\,7 perform concurrently to find out whether the augmentation of the T number will strengthen the coverage. Tx\,7 is set above the rear seat, close to the ceiling of the vehicle.
In the final stage, simulations involving more than two Txs are conducted. The cases are listed in table~\ref{tab:ratem}. Tx~5 and Tx~6 are at the front corner of the vehicle. Tx~8 and Tx~9 are at the rear corner of the vehicle. The corresponding results are illustrated using both coverage maps and coverage-threshold curves, allowing for a detailed comparison of the coverage performance as a function of the number of Ts. In these scenarios, the Ts are strategically placed at multiple locations throughout the vehicle interior.

Optimization is carried out with a two-staged strategy.
which aims to maximize system throughput by selecting optimal Tx numbers and coordinates. First, the best quantity of Tx is determined by analyzing the coverage and capacity of single and multiple cases. In the second stage, an optimizer is applied to find the optimal coordinates. Therefore, the proposed optimization framework can be mathematically expressed as
\begin{subequations}
\label{opt_problem}
\begin{align}
    \mathop{\mathrm{max}}\limits_{\boldsymbol{x},\boldsymbol{y},\boldsymbol{z}} ~\mathop{\mathrm{max}}\limits_{N}    & ~\frac{1}{N}\sum_{i}\mathbb{E}[\mathcal{R}(x_i,y_i,z_i)] \label{subequatios:opt} \\ 
    s.t. & ~ (x_i,y_i,z_i) \in V \label{subequation:V} \\ 
    &~ 1 \leq N \label{subequation:N}\\
    &~ 1 \leq i \leq N \label{subequation:i} \\
    &~ P_c(\gamma_{i})>P_{th} \label{subequation:Pc}
\end{align}
\end{subequations}
where $(x_i,y_i,z_i)$ stands for the $i^{th}$ Tx coordinates, and all the coordinates are constrained in $V$, i.e., the interior space of the vehicle, as described in~(\ref{subequation:V}). In (\ref{subequation:N}) and (\ref{subequation:i}), $N$ is the quantity of Txs, ensuring $i$ is between $1$ and $N$. $\gamma_{i}$ and $P_{th}$ represent the desired threshold for each Tx and the expected coverage probability, which is denoted in~(\ref{subequation:Pc}).

\section{Data Post-Processing and Hybrid Channel Modeling}
\label{section:Channelmodel}
An in-vehicle three-dimensional channel model is established based on the confined space characteristics and material properties of the vehicle interior.

\subsubsection{Signal Model}
The received signal model in the V2X channel can be expressed by the CIR as in
\begin{equation}
\begin{split}
    h(\tau, \theta, \varphi) & = \sum_{\ell=1}^{L} \alpha_{\ell} \delta(\tau - \tau_{\ell}) \cdot \delta(\theta - \theta_{\ell}) \cdot\delta(\varphi - \varphi_{\ell}),
\end{split}
\label{equation: CIR}
\end{equation}
where $L$ is the total number of propagation paths, and $\alpha_{\ell}$, $\tau_\ell$, $\theta_{\ell}$, and $\varphi_{\ell}$ represent the complex channel coefficient, the delay, the zenith angle, and azimuth angle of the $\ell$-th propagation path.

Through the measurement experiment, CFR $H(f, \theta, \varphi)$ is obtained, where $f$ is the carrier frequency. The CIR can be obtained through the inverse discrete Fourier transform (IDFT).
Thus, the power $P(\tau, \varphi)$ can be derived as
\begin{equation}
P(\tau, \varphi) = \sum_{\theta} |h(\tau, \theta, \varphi)|^2 ,
\label{equation: PADP}
\end{equation}
According to (\ref{equation: PADP}), the relationship between power and delay or angle is explicit. Thus, a PADP diagram can be plotted to depict the power of different paths.

To estimate channel parameters from measurement CIR, a local maxima search method is used to extract MPC parameters from the PADP results~\cite{7481455}. 
The estimated paths after the local maxima search algorithm can be presented as
\begin{equation}
\mathcal{P}=\{ \tau_{\ell}, \theta_{\ell}, \varphi_{\ell}, P_{\ell} \}_{\ell=1}^{L}.
\label{equation: Path}
\end{equation}

\subsubsection{Material Properties}
\label{section: MP}
In the V2X model, when THz waves encounter the surfaces of various materials inside the vehicle, both reflection and transmission occur~\cite{8674541}. According to the reflection theory, the reflected power depends on the material properties and the material thickness. In the simulation process, the thickness of the material is abstracted as an attribute value, based on which the reflection loss is calculated~\cite{ITU-R-P2040-3}. When the reflection object has a finite thickness, i.e., $d$, the reflection coefficient can be acquired by
\begin{equation}
R = \frac{R'(1-e^{-\mathrm{j}2q})}{1-(R')^2e^{-\mathrm{j}2q}},
\label{equation: RL}
\end{equation}
where $R'$ is the reflection coefficient depending on the polarization of the incident electric field, and
\begin{equation}
q=\frac{2\pi d}{\lambda}\sqrt{\eta-\sin^2\theta_0},
\end{equation}
which is related to the thickness of the material. $\eta$ and $\theta_0$ stand for the complex relative permittivity and the angle of incidence. Reflection loss is defined as the reflection coefficient in decibel form and can be derived from measured power.
\begin{align}
RL_{\ell}[\mathrm{dB}] & =PL_{\ell}[\mathrm{dB}]-FSPL_{\ell}[\mathrm{dB}]\\
&=20\log_{10}(P_{omni}(\tau_{\ell}))-20\log_{10}(4\pi f\tau_{\ell}), \notag
\end{align}
where RL, PL, and FSPL respectively stand for reflection loss, path loss, and free space path loss. $P_{omni}(\tau)=\mathop{\mathrm{max}}\limits_{\theta,\varphi}|h(\tau,\theta,\varphi)|^2$ represents the omnidirectional PDP.
A THz-TDS-based measurement system is utilized to derive material properties~\cite{pimrc}. Reflection loss of common in-vehicle materials, including glass, metal, and leather, is calculated to be $2.42$~dB, $9.45$~dB, $17.75$~dB, and $20.11$~dB, respectively. Reflection loss can be calculated from measurement results and validated by the material database obtained by THz-TDS.

\subsubsection{Hybrid Channel Model}
A hybrid channel model is established, combining RT results and measurement data, which can be expressed as
\begin{equation}
h_{hybrid}(\tau,\theta,\varphi)=h_{RT}(\tau,\theta,\varphi)+h_{s}(\tau,\theta,\varphi),
\label{equation: hybrid}
\end{equation}
where $h_{RT}$ represents the CIR of RT results, and $h_{s}$ stands for the statistical model of measurement results, which can be calculated as the aggregation of CIR of subpaths in each cluster
\begin{align}
    h_{s}(\tau,\theta,\varphi)=&\sum_{\ell}^{L}\sum_{p}\alpha_{\ell,p} \delta(\tau - \tau_{\ell,p}) \cdot \delta(\theta - \theta_{\ell,p}) \cdot\delta(\varphi - \varphi_{\ell,p}) \notag\\
    +& \sum_{q}^{L_s}\sum_{s}\alpha_{q,s} \delta(\tau - \tau_{q,s}) \cdot \delta(\theta - \theta_{q,s}) \cdot\delta(\varphi - \varphi_{q,s}),
\end{align}
where $p$ denotes the $p^{th}$ subpath of ray-traced $\ell^{th}$ cluster, while $q$ stands for the $q^{th}$ subpath of non-RT $s^{th}$ cluster. $L$ and $L_s$ represent the number of ray-traced clusters and non-RT cluster.

\section{Case Studies}
\label{section:validation}
In this section, the hybrid channel model of three cases demonstrated in table~\ref{tab:cfv} are established. It should be noted that, compared to the previous work~\cite{9466322}, the material properties are especially considered to map the hybrid channel model with the material database. The difference and analysis between with-human and without-human, along with with-window and without window, will be concluded.
\subsection{Case A. Without Human/Window-on}

\begin{figure}[t]
\centering
    \includegraphics[width=0.8\columnwidth]{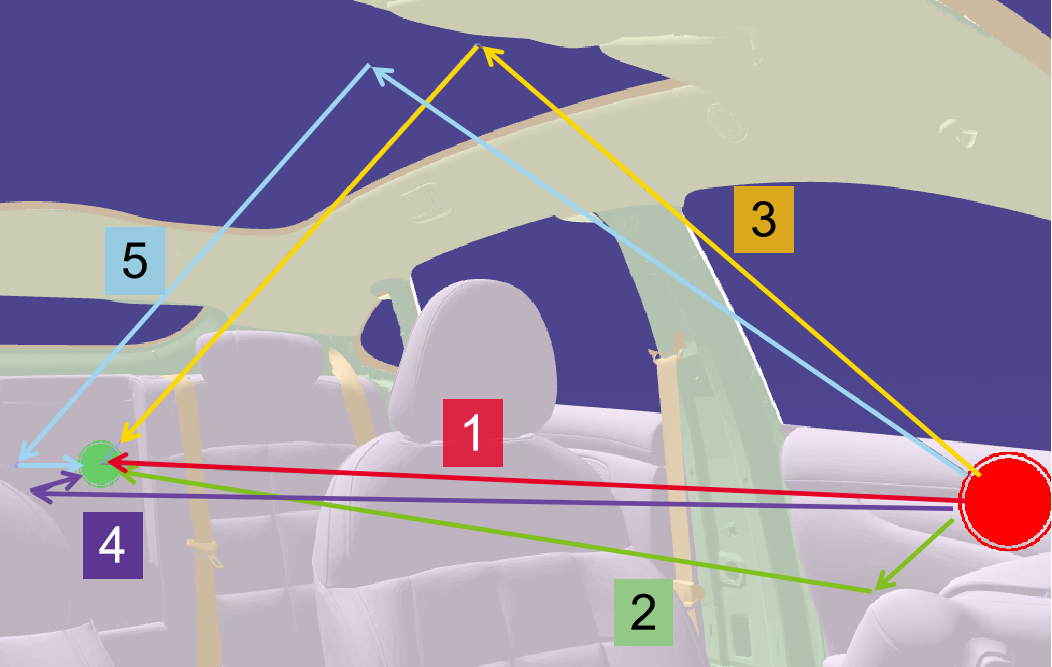}
\caption{Simulation results for exemplary LoS case, i.e., Rx placing above the middle rear seat.}
\label{fig:Multipath}
\end{figure}
\begin{figure}[t]
\centering
\subfloat[]{
    \hspace{0.03\columnwidth}
    \includegraphics[width=0.9\columnwidth]{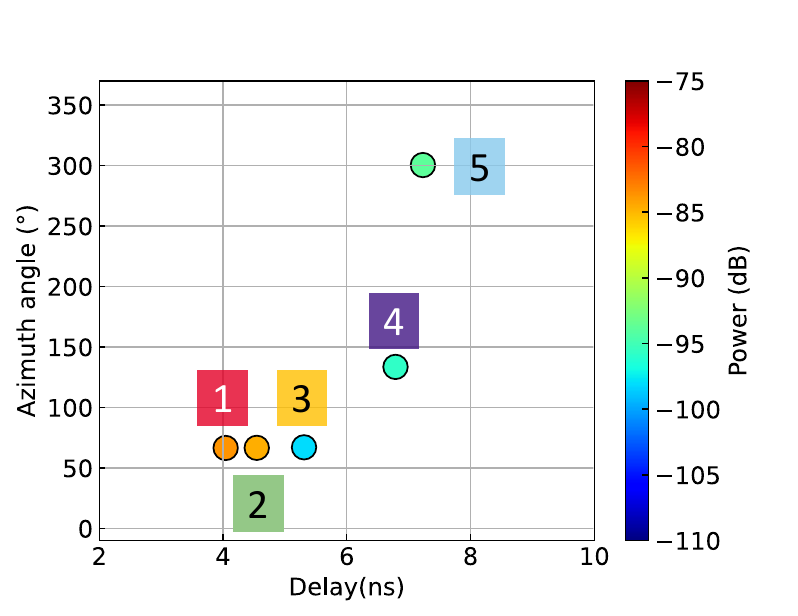}
}
\hfill
\subfloat[]{
    \includegraphics[width=0.85\columnwidth]{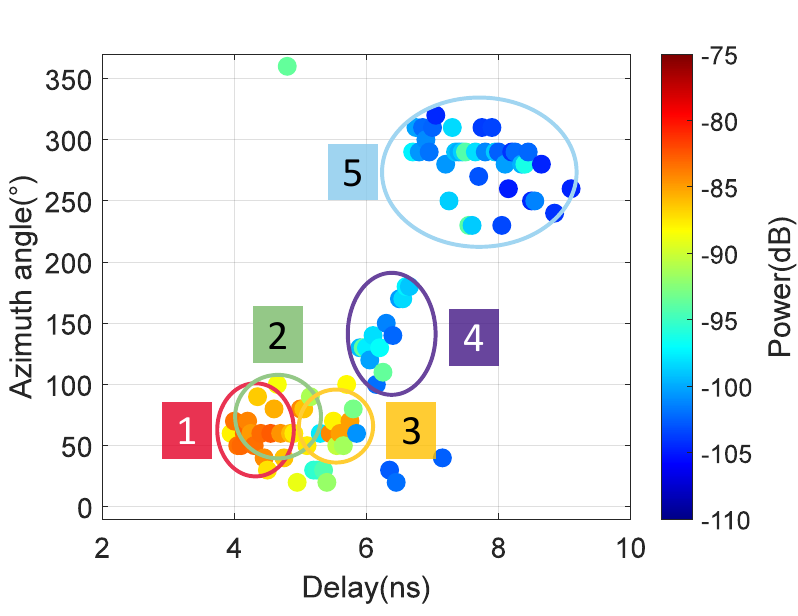}
}
\caption{Exemplary PADP at Rx~2. (a) Simulated, (b) Measurement.}
\label{fig:PADP}
\end{figure}

\begin{figure}[t]
\centering
\subfloat[]{
    \includegraphics[width=0.9\columnwidth]{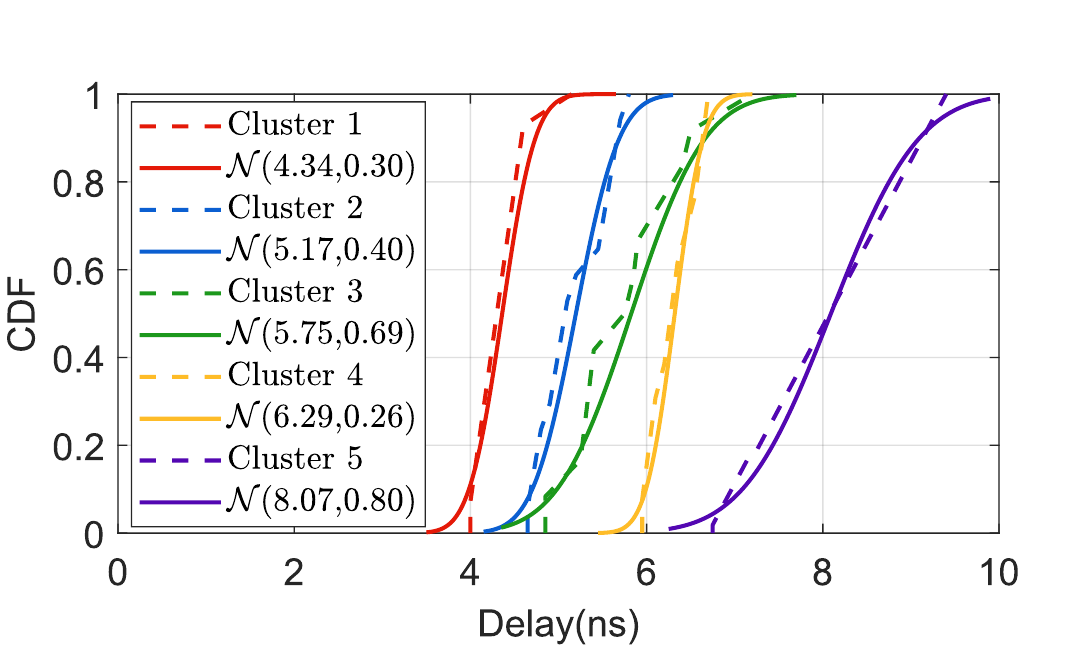}
}
\hfill
\subfloat[]{
    \includegraphics[width=0.9\columnwidth]{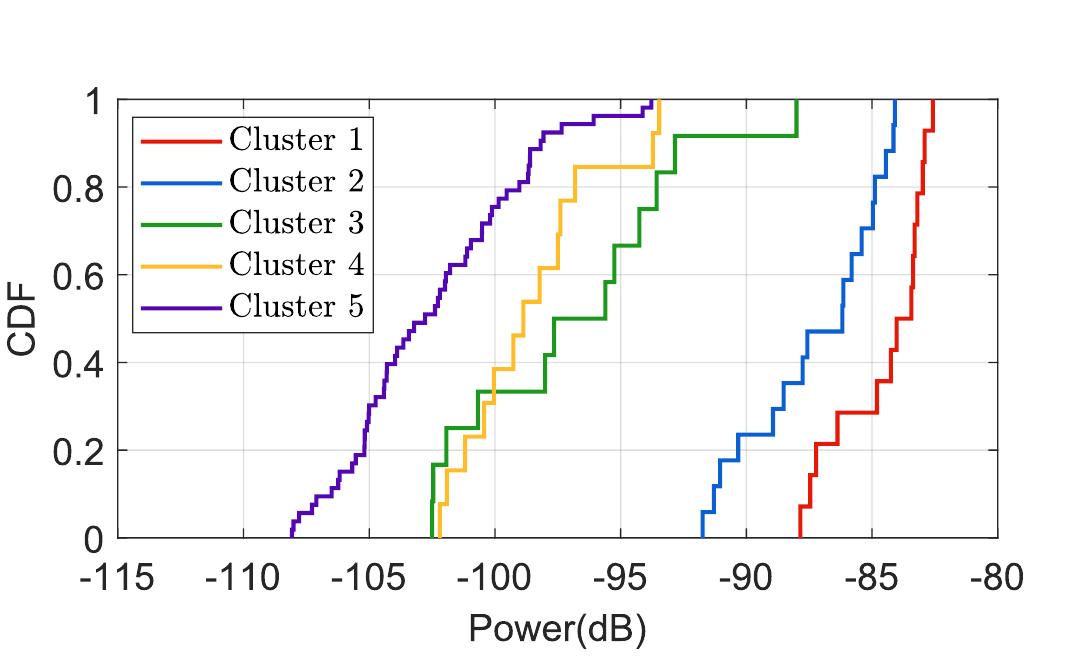}
}
\caption{CDF for each cluster at Rx~2. (a) Delay distribution (b) Power distribution}
\label{fig:CDF}
\end{figure}

\begin{table*}[]
    \caption{Detailed delay, angle and power of simulation and measurement}
    \label{tab:deda}
    \centering
    \begin{tabular}{ccccccccc}
    \toprule
        \multirow{2}*{Path} & \multicolumn{2}{c}{Delay/$\mathrm{ns}$} & \multicolumn{2}{c}{Azimuth/$^{\circ}$} &  \multicolumn{2}{c}{Zenith/$^{\circ}$}  &  \multicolumn{2}{c}{Power/$\mathrm{dB}$}\\
        \cline{2-9}
        ~ &  Simulation  & Measurement  &  Simulation  & Measurement  &  Simulation  & Measurement  &  Simulation  & Measurement  \\
    \midrule
         1 & 4.04 & 4.15 & 66.51 & 60 & 0.41 & 0 & -83.66 & -82.91 \\
         2 & 4.55 & 4.2 & 66.62 & 70 & 0.36 & 0 & -84.68 & -84.02 \\
         3 & 5.31 & 5.3 & 67.06 & 60 & 35.06 & 30 & -97.17 & 97.64 \\
         4 & 6.79 & 6.4 & 133.54 & 140 & 35.25 & 30 & -101.24 & -102.193 \\
         5 & 7.23 & 7.5 & 300.39 & 290 & 2.13 & 0 & -93.84 & -94.13 \\
    \bottomrule
    \end{tabular}
\end{table*}

Simulation experiments with all the different Rx positions are conducted. Both LoS and obstructed-LoS (OLoS) cases are illustrated. For example, in the Tx~1 case, the traced rays for Rx~1 (i.e., OLoS case) and Rx~2 (i.e., LoS case) are depicted in Fig.~\ref{fig:Multipath}, and the exemplary simulated power-delay-angle profile (PADP) of case Rx~2 is illustrated in Fig.~\ref{fig:PADP}~(a). 

Meanwhile, the measurement data are processed, and the results are shown in Fig.~\ref{fig:PADP}~(b). Due to the wide measurement bandwidth ($20$~GHz), MPCs can be clearly distinguished in the delay domain. The local maxima search method mentioned in section~\ref{section:Channelmodel} is used to extract MPC parameters from the PADP results. The measurement reveals more MPCs than the RT simulation, primarily due to the presence of rough surfaces inside the vehicle that introduce diffuse reflections, an effect not considered by the simulation. Comparing the simulation and measurement results, path~$1$ corresponds to the LoS path, which undergoes no obstruction or reflection. It exhibits the strongest power at $-83.66$~dB and the shortest delay of $4.04$~ns. Both its azimuth and zenith angles closely match between simulation and measurement. Detailed results, including zenith data, are summarized in Table~\ref{tab:deda}. Paths~$2$-$5$ represent NLoS components, resulting from single-bounce reflections, most commonly off window glass surfaces, which are identifiable in the measured data. Other Rx positions exhibit similar trends, with delay errors within $0.4$~ns, azimuth errors within $10.5^{\circ}$, zenith errors within $6.5^{\circ}$, and power differences within $1.5$~dB. The primary discrepancies between simulation and measurement are attributed to minor geometry mismatches of the transceiver setup. Overall, the simulation results align well with measurements, validating the employment of the simulation for channel characterization and wireless planning in vehicular scenarios.

A hybrid channel model is established, combining RT results and measurement data. CDF of delay is illustrated in Fig.~\ref{fig:CDF}.
The CDF curves are used to construct a statistical channel model, in which MPCs generated by diffuse scattering are grouped into clusters that correspond one-to-one with deterministic paths.
The fitted CDF curves of cluster $1-5$ have mean values of $4.34$~ns, $5.17$~ns, $5.75$~ns, $6.29$~ns, and $8.07$~ns, representing the statistical delay of the deterministic components.

As shown in Fig.~\ref{fig:CDF}, the power CDFs of different clusters extracted from the measurement results at the Rx2 position are illustrated. Each cluster corresponds to a multipath component obtained from the deterministic channel modeling, while the random components within each cluster are represented by the CDF, serving as the statistical channel modeling results. The deterministic model further identifies the reflecting materials associated with each multipath component, establishing a one-to-one correspondence with the statistical results. For instance, Cluster~2, corresponding to a single reflection from the car window, exhibits a higher-power CDF curve and fewer random components, owing to the smooth surface of the window. In contrast, Cluster~5, resulting from two reflections of the window and the seat, shows a lower-power CDF curve. This indicates higher reflection loss due to multiple reflections, as well as greater path loss caused by the increased propagation distance. Furthermore, the larger number of random components within this cluster suggests that the rough surface of the car seat introduces significant diffuse scattering.

\begin{figure}[t]
\centering
\includegraphics[width=0.9\columnwidth]{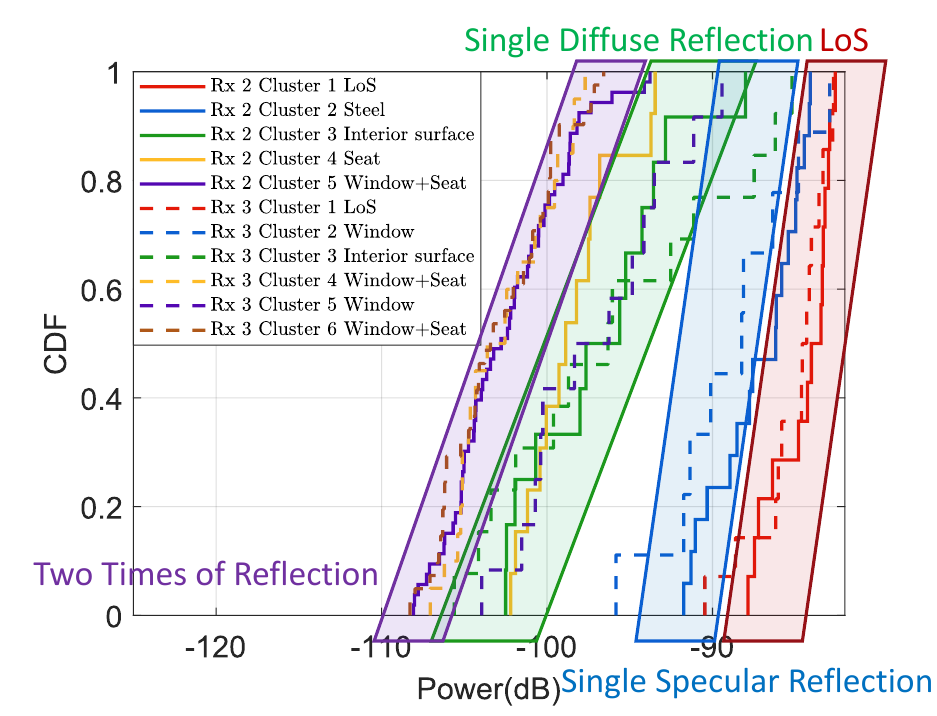}
\caption{CDF of power of different clusters among Rxs assigned by reflection material.}
\label{fig:CDFs}
\end{figure}

The power CDFs of clusters measured at different Rx locations can be classified according to their power values, and these different categories correspond one-to-one with the reflection paths. 
Fig.~\ref{fig:CDFs} illustrates the CDFs of multipath power at different Rx positions. The characteristics of the reflecting surfaces give rise to distinct distributions of random MPCs. Specifically, a single reflection from the glass results in relatively higher received power with fewer random components, whereas clusters associated with two reflections from the window and the seat exhibit lower received power accompanied by a larger number of random components. Furthermore, the distribution patterns remain largely consistent across different Rx positions, demonstrating that the power CDFs effectively capture the material properties of the propagation paths and align well with the deterministic modeling results. 
The average power is compared with the reflection loss mentioned in section~\ref{section: MP}, and the results are listed in table~\ref{tab:tds}. The calculated reflection losses differ from those in the material database by no more than $3$~dB, enabling accurate material identification and thus allowing extraction of material information from the cluster-based hybrid model.

\begin{table}[t]
\caption{RL of different clusters and the material identified.}
\label{tab:tds}
\centering
\begin{tabular}{ccp{0.18\columnwidth}|ccp{0.18\columnwidth}}
\toprule
Cluster & RL\,/\,dB & Identified material & Cluster & RL\,/\,dB & Identified material \\
\midrule
Rx~2-2 & 3.80 & Steel & Rx~3-2 & 5.32 & Glass\\
Rx~2-3 & 14.62 & Rubber & Rx~3-3 & 13.51 & Rubber\\
Rx~2-4 & 16.80 & Rubber & Rx~3-4 & 20.73 & Glass+Rubber\\
Rx~2-5 & 22.70 & Glass+Rubber & Rx~3-5 & 16.56 & Rubber\\
 &  &  & Rx~3-6 & 23.54 & Glass+Rubber\\
\bottomrule
\end{tabular}
\end{table}

\subsection{Case B. Without Human}

\begin{figure*}[t]
\centering
\subfloat[]{
    \includegraphics[width=0.66\columnwidth]{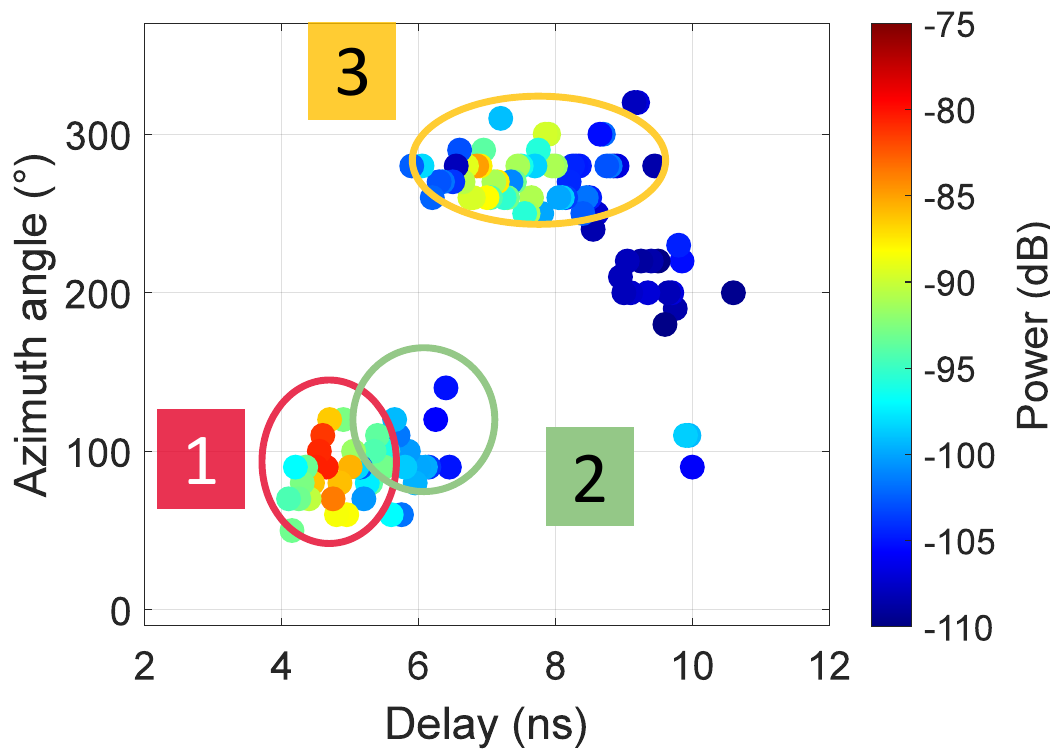}
}
\subfloat[]{
    \includegraphics[width=0.66\columnwidth]{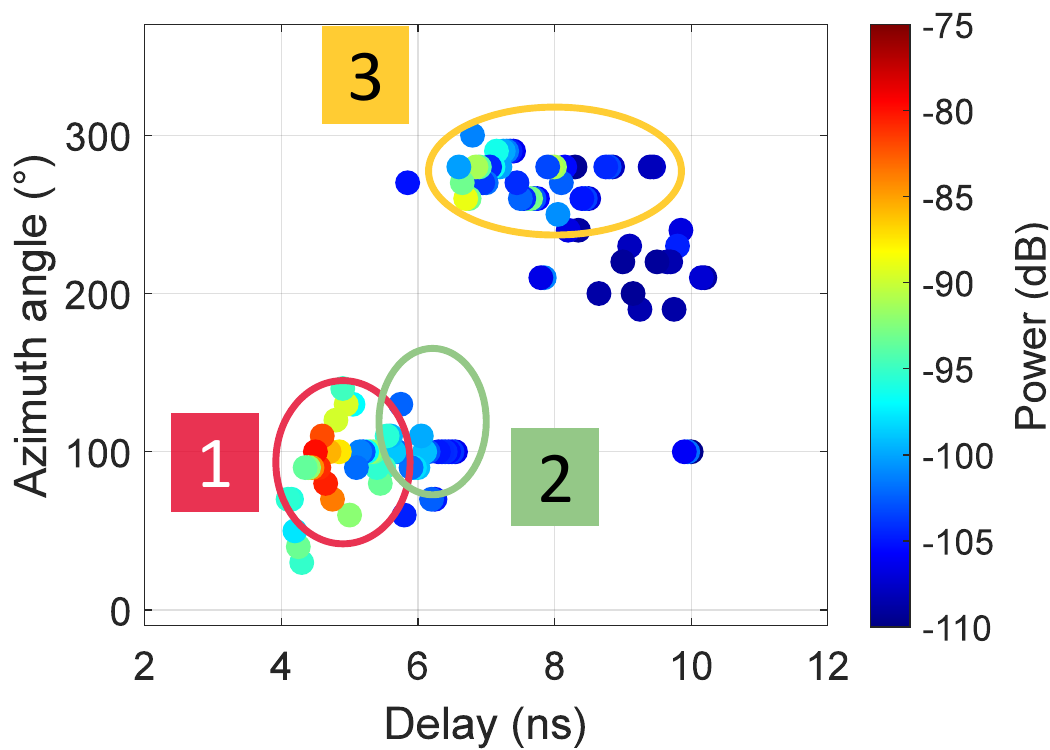}
}
\subfloat[]{
    \includegraphics[width=0.66\columnwidth]{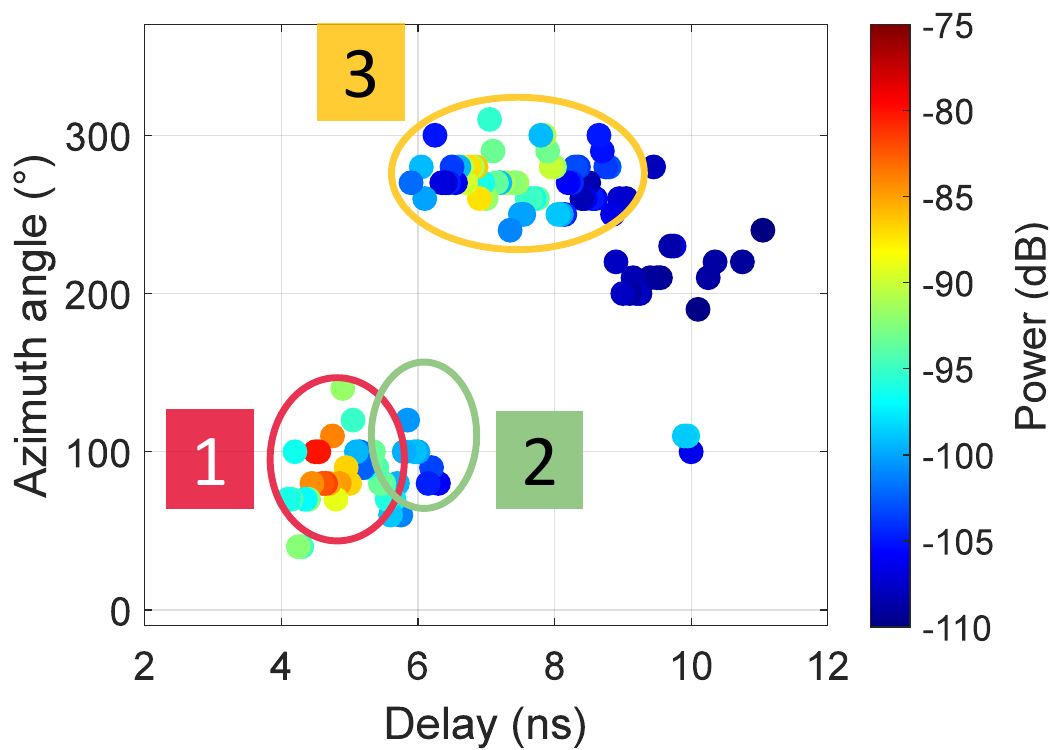}
}
\hfill
\hspace{0.02\columnwidth}
\subfloat[]{
    \includegraphics[width=0.66\columnwidth]{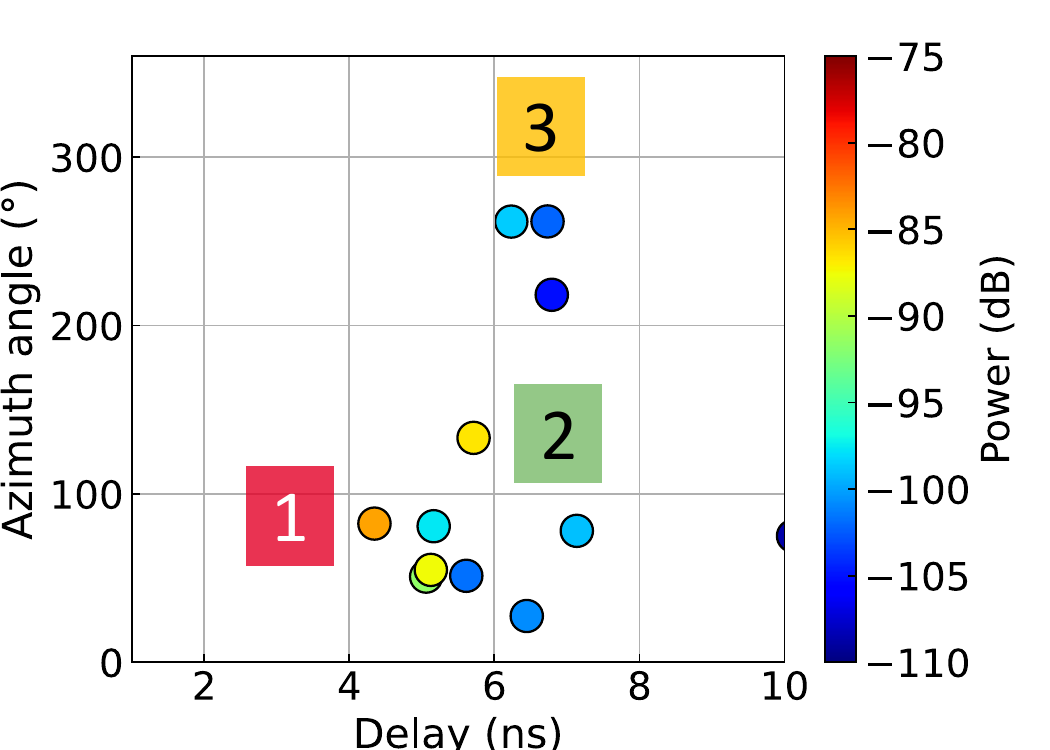}
}
\subfloat[]{
    \includegraphics[width=0.66\columnwidth]{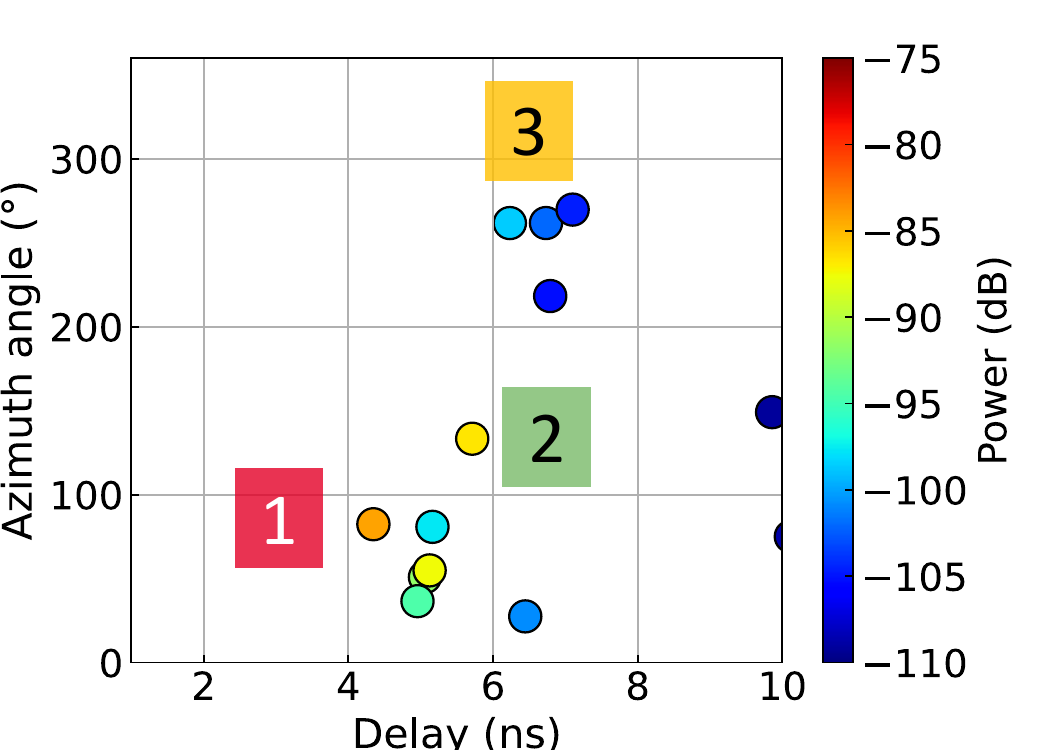}
}
\subfloat[]{
    \includegraphics[width=0.66\columnwidth]{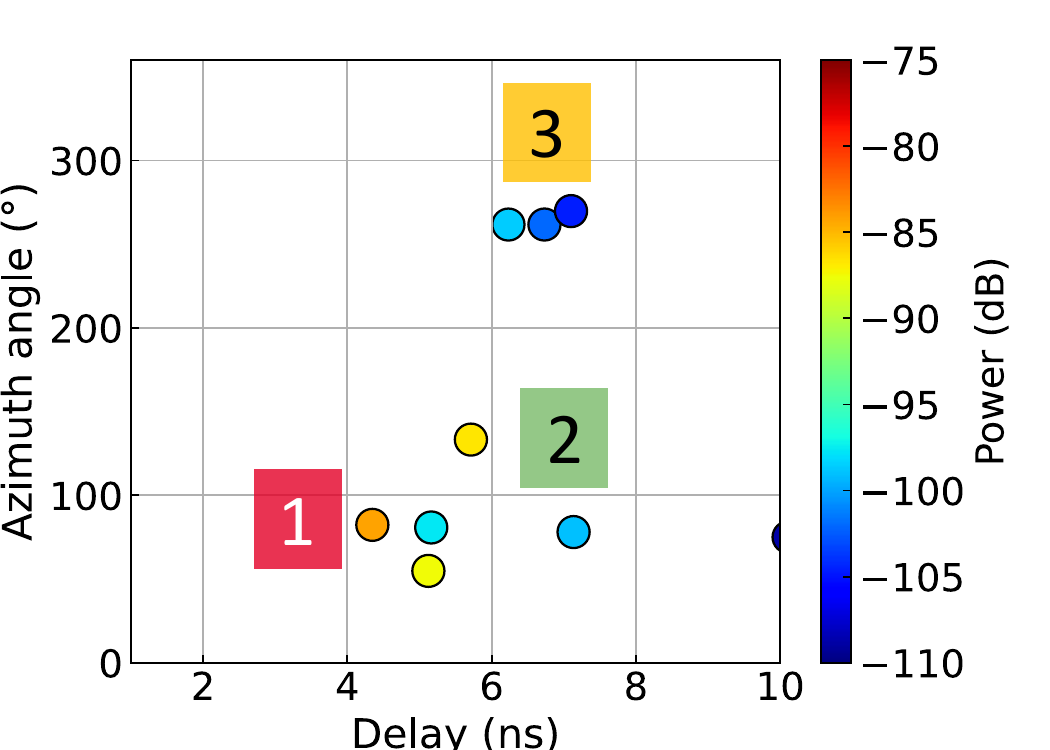}
}
\caption{Exemplary RT Simulated PADP at Rx~1 with human. (a) Simulated, without human, (b) Simulated, with human, (c) Simulated, without window, (d) Measurement, without human, (e) Measurement, with human, (f) Measurement, without window.}
\label{fig:WH}
\end{figure*}


Simulations under the circumstances of a human inside the vehicle are conducted. Fig.~\ref{fig:WH} (b) and (e) depict the exemplary measured and simulated PADP of the OLoS case Rx~1. The MPCs in the simulation data and the measurement data can still be matched one-to-one. From both the simulated and measured PADP figures, three distinct MPCs can be clearly observed. Among them, Path~1 corresponds to the obstructed LoS component, while Paths~2 and 3 represent reflected components. 
The power of OLoS in the measurement is $-82.91$~dB, while the simulated power of OLoS is $-84.29$~dB.
Similarly, due to the presence of diffuse scattering, a large number of diffuse MPCs are also generated. The angular error is within 10 degrees, the delay error is within 0.7~ns, and the power error is within $5$~dB. Even in the presence of a human body blockage, the simulation results still correspond well with the measurement data.

\begin{figure}[t]
\centering
\subfloat[]{
    \includegraphics[width=0.9\columnwidth]{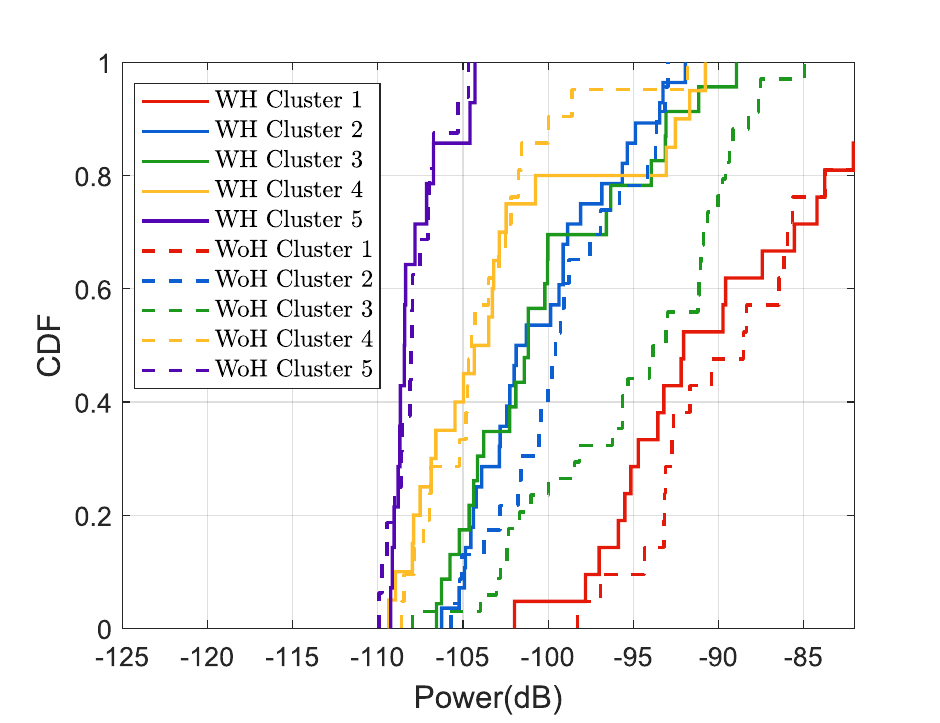}
}
\hfill
\subfloat[]{
    \includegraphics[width=0.9\columnwidth]{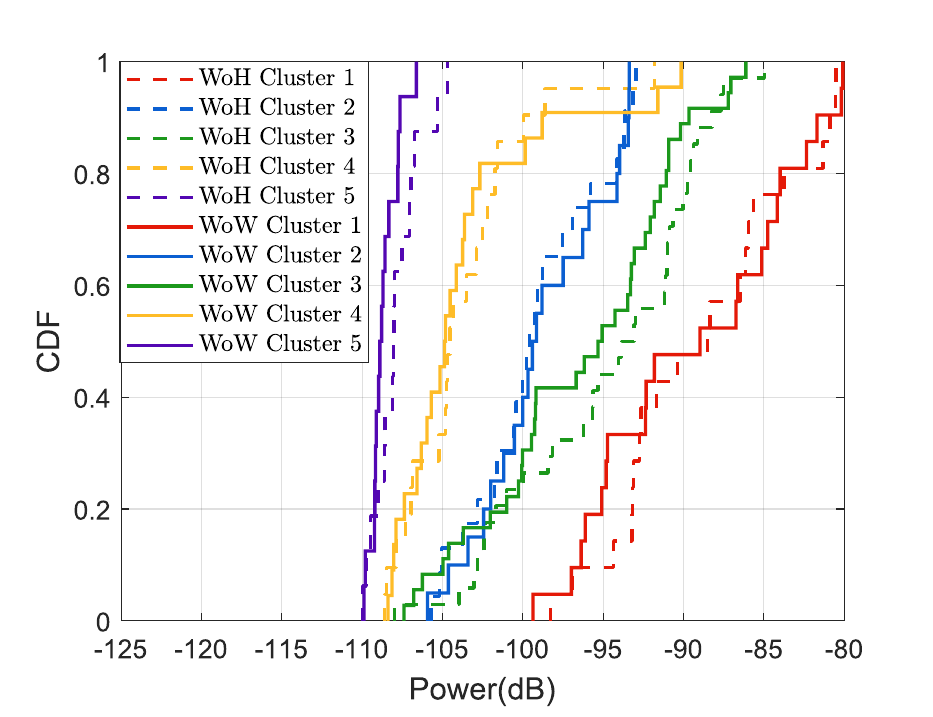}
}
\caption{CDF for each cluster at Rx~1. compared with the empty-vehicle case (a) Human-occupied (b) Window-off.}
\label{fig:CDFw}
\end{figure}

By comparing Fig.~\ref{fig:WH} (b) (i.e., the case with a human presence inside the vehicle) and Fig.~\ref{fig:WH} (a) (i.e., the case without a human), it is observed that although the human body causes blockage, it only affects a subset of the MPCs, introducing a penetration loss of approximately 10~dB. The major affected range is the left rear seat. As shown in Fig.~\ref{fig:CDFw}~(a), the statistical modeling results are compared between the empty-vehicle and occupied-vehicle scenarios. The presence of passengers introduces interference to certain MPCs and reduces the average power of the OLoS path by approximately 10 dB.

\subsection{Case C. Without Window}
The measured and simulated PADP of case Rx~1 with the vehicle's window open and without a human is demonstrated in Fig.~\ref{fig:WH} (c) and (f). From the comparative analysis between simulation and measurement results under the no-window scenario, all three dominant MPCs can be clearly matched one-to-one. Their power, angle, and delay are all within acceptable error margins, validating that the DT approach can reliably simulate the scenario where the car windows are rolled down.
In addition, the measurement results without windows can be compared with those obtained with windows, which is shown in Fig.~\ref{fig:WH} (a). The simulation results show that fewer reflective MPCs originate from the vehicle’s side windows. As shown in Fig.~\ref{fig:CDFw}~(b), the statistical modeling results are compared between the scenarios with and without side windows. Different clusters exhibit a high degree of consistency, indicating that in this scenario, almost no MPCs are generated through reflections from the side windows. Therefore, the measurement results indicate that whether the windows are open or closed has little impact on the power of individual MPCs.

\section{Wireless Planning based on Digital Twin}
\label{section:WP}
In this section, the results of wireless planning will be demonstrated. Firstly, a single-Tx case is discussed, where the coverage map is established among the Txs, and the optimal Tx is determined. Moreover, the multi-Tx case is examined to analyze the interference, coverage, and system rate with multiple Txs.

\subsection{Case A. Single Transmitter}
In this case, Tx~1 to 4 is set one by one, playing the role of the single AP, and $80$ Rxs are placed inside the vehicle, representing UEs. Most of the Rxs are distributed in areas with a higher probability of actual presence, such as above the seats.
\subsubsection{Coverage Probability}
\begin{figure}[t]
\centering

\subfloat[]{
    \includesvg[width=0.9\columnwidth]{figures/WP07.svg}
}
\hfill
\subfloat[]{
    \includesvg[width=0.9\columnwidth]{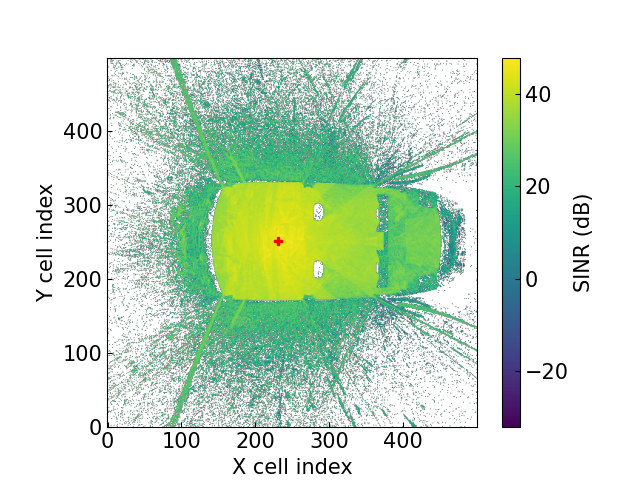}
}
\caption{Coverage of different Tx. (a) Coverage probability with threshold (b) Coverage map at z = $1.0$m.}
\label{fig:CP}
\end{figure}

\begin{figure}[t]
\centering
\includesvg[width=0.9\columnwidth]{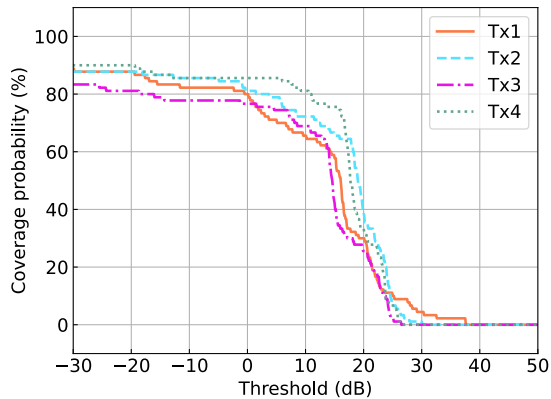}
\caption{SINR of different Tx.}
\label{fig:MaxR}
\end{figure}


The 2D distance between the $i^{\mathrm{th}}$ Rx and the Tx is denoted as $u_{i}$, while the LoS 3D LoS distance is denoted as $v_{i0}$, where $i$ stands for the index number of the Rx. Meanwhile, the NLoS distance is denoted as $v_{ij}$, where $j$ stands for the $j^{\mathrm{th}}$ path. 
To better analyze the channel and depict interference and coverage in a network, a statistical model of the THz signal can be described through path loss as
\begin{equation}
L(u_0) = K(u_0)g(u_0),
\label{equation: PL}
\end{equation}
where $K(u_0)$ corresponds to free space path loss (FSPL) and the molecular absorption with the distance between the target Rx and the Tx, which describes the large-scale loss. And $g(u_0)$ describes the small-scale fading path loss, which follows a distribution with distance-dependent parameters.
Interference can be defined as the received power from other Txs plus noise, which can be characterized numerically by SINR~\cite{8620216}.
SINR discloses the transmission quality of a channel, which can be calculated as
\begin{equation}
\begin{split}
    \mathrm{SINR} = \frac{S}{I + P_N} = \frac{P_tG_0K_ug(u)}{\sum_{i\in{\Phi_0/AP_0}}P_tG_iK_{x_i}g(x_{i}) + N_0B},
\end{split}
\label{equation: SINR}
\end{equation}
where $P_t$ stands for the transmitted power of Tx, $G_0$ and $G_i$ denote the gain power of Tx and the target Rx and the $i^{th}$ other Tx or potential eavesdroppers, respectively, $N_0$ refers to the noise power spectral density, and $B$ means the bandwidth.





In the single-Tx case, there are no other APs, and SINR can be simplified as
\begin{equation}
\begin{split}
    \mathrm{SINR(dB)} = P_r(\mathrm{dB})-P_N(\mathrm{dB}).
\end{split}
\end{equation}
As a result, the SINR is primarily determined by the position of the Ts and Rxs.

Coverage probability represents the likelihood that the SINR exceeds a given threshold ($T$)~\cite{8620216}, which can be computed as
\begin{equation}
\begin{split}
    P_c(T)=\iiint_{V}(p_L(u)P_{c,L}(T)+p_N(u)P_{c,N}(T))f(v)\mathrm{d}V,
\end{split}
\label{equation: CP}
\end{equation}
where $f(v)$ is the probability distribution of the Rxs, and $P_{c,L}$ is the conditional coverage probability where the 3D distance is given, and can be calculated as
\begin{equation}
\begin{split}
    P_{c,L}(T) = \mathbb{P}(\mathrm{SINR}>T|u)=\mathbb{P}(\frac{P_tG_0K_ug(u)}{N_0B}>T).
\end{split}
\end{equation}

In single-Tx case, the SINR primarily corresponds with the position and distance between Tx and Rx. Therefore, the coverage map can be obtained by calculating the receive power between Tx and each Rx and checking whether it is larger than the threshold.

Fig.~\ref{fig:CP} (a) illustrates the variation of coverage probability with respect to the SINR threshold. Different Txs are represented by lines with distinct colors and styles. At lower thresholds, the coverage probability remains high for all Tx options. However, as the threshold increases, a noticeable drop occurs around $19$~dB, indicating a critical point for estimating the system's effective coverage. Under the commonly used threshold of $10$~dB, Tx~4 achieves the highest coverage.

Fig.~\ref{fig:CP} (b) presents a top-view coverage map. This figure illustrates the relationship between SINR and planar coordinates at a height of $z=1\mathrm{m}$, with the Tx positioned at Tx~4. The region closest to the Tx (marked by the red cross at the center) exhibits the highest SINR values, while the outer edges of the vehicle experience signal blockage, resulting in zero received power. Due to obstructions, the SINR in the rear seats and trunk area is noticeably reduced. Overall, under the single-Tx scenario, SINR values exceeding $20$~dB can be observed in the area spanning the front and rear seats, and the signal in the trunk region does not completely vanish, indicating strong transmission capability.

\subsubsection{System Rate and Optimal Tx Position}
Due to the constraints imposed by the vehicle's shape, the optimal Tx location cannot be placed inside the materials or outside the vehicle. Therefore, a two-staged strategy is proposed to determine the optimal Tx position. First, a set of candidate Tx locations is predefined based on the vehicle structure and practical deployment feasibility. The coverage probability and system data rate are then evaluated for each candidate to identify the optimal location, functioning as the initial location $(x_0,y_0,z_0)$ of the optimizer. Subsequently, optimization techniques are employed to compute the Tx position, which is then validated to assess its practical applicability. 

As is proposed in (\ref{opt_problem}), SINR, coverage, and data rate are jointly considered to determine the optimal Tx position. The system rate is the maximum possible transmission rate in the V2X system. The average rate can be derived as
\begin{equation}
\begin{split}
    \mathbb{E}[\mathcal{R}] & =  \frac{B}{N} \, \mathbb{E} [\mathrm{log_2}(1+\mathrm{SINR})]=
      \frac{B }{N\,\mathrm{ln}\,2}\int_0^{\infty}\frac{P_{c}(t)}{1+t}\mathrm{d}t,
\end{split}
\end{equation}
where $N$ stands for the number of APs in the vehicle, and $\mathrm{SINR}_i$ stands for the $i^{\mathrm{th}}$ UE with the $i^{\mathrm{th}}$ AP. The average rate is closely related to the coverage probability. A wider coverage area inside the vehicle can enhance the maximum communication rate of the V2X system, thereby improving the overall communication quality.
Therefore, the optimization of the single-Tx case can be expressed as
\begin{align}
    \mathop{\mathrm{max}}\limits_{x_o,y_o,z_o} ~    & ~\frac{B }{N\,\mathrm{ln}\,2}\int_0^{\infty}\frac{P_{c}(t|x_o,y_o,z_o)}{1+t}\mathrm{d}t \\ 
    s.t. & ~ (x_o,y_o,z_o) \in V \notag \\ 
    &~ P_c(\gamma_{o})>P_{th}. \notag
\label{equation:opt1}
\end{align}

The SINR performance of different Txs is illustrated in Fig.~\ref{fig:MaxR}, where each Tx is represented using distinct colors and shapes. Missing data points indicate scenarios where no valid MPCs exist between a specific Tx and Rx. The x-axis represents $30$ different Rx locations, covering both LoS and NLoS positions. In most cases, Tx~2 exhibits a higher SINR compared to the other Txs. However, for certain Rxs, no available paths are observed for Tx~1-3, while Tx~4 maintains connectivity with all Rx positions. The transmission rates of Tx~1-4 are $128.23$~Gbps, $146.71$~Gbps, $114.77$~Gbps, and $150.78$~Gbps. Among all cases, Tx~4 achieves the highest rate. The analysis indicates that Tx~4, positioned beneath the ceiling of the car, provides the widest coverage and the highest transmission rate, making it the most likely candidate for the optimal Tx placement.

\begin{figure}[t]
\centering
\subfloat[]{
    \includesvg[width=0.9\columnwidth]{figures/WP16.svg}
}
\hfill
\subfloat[]{
    \includegraphics[width=0.8\columnwidth]{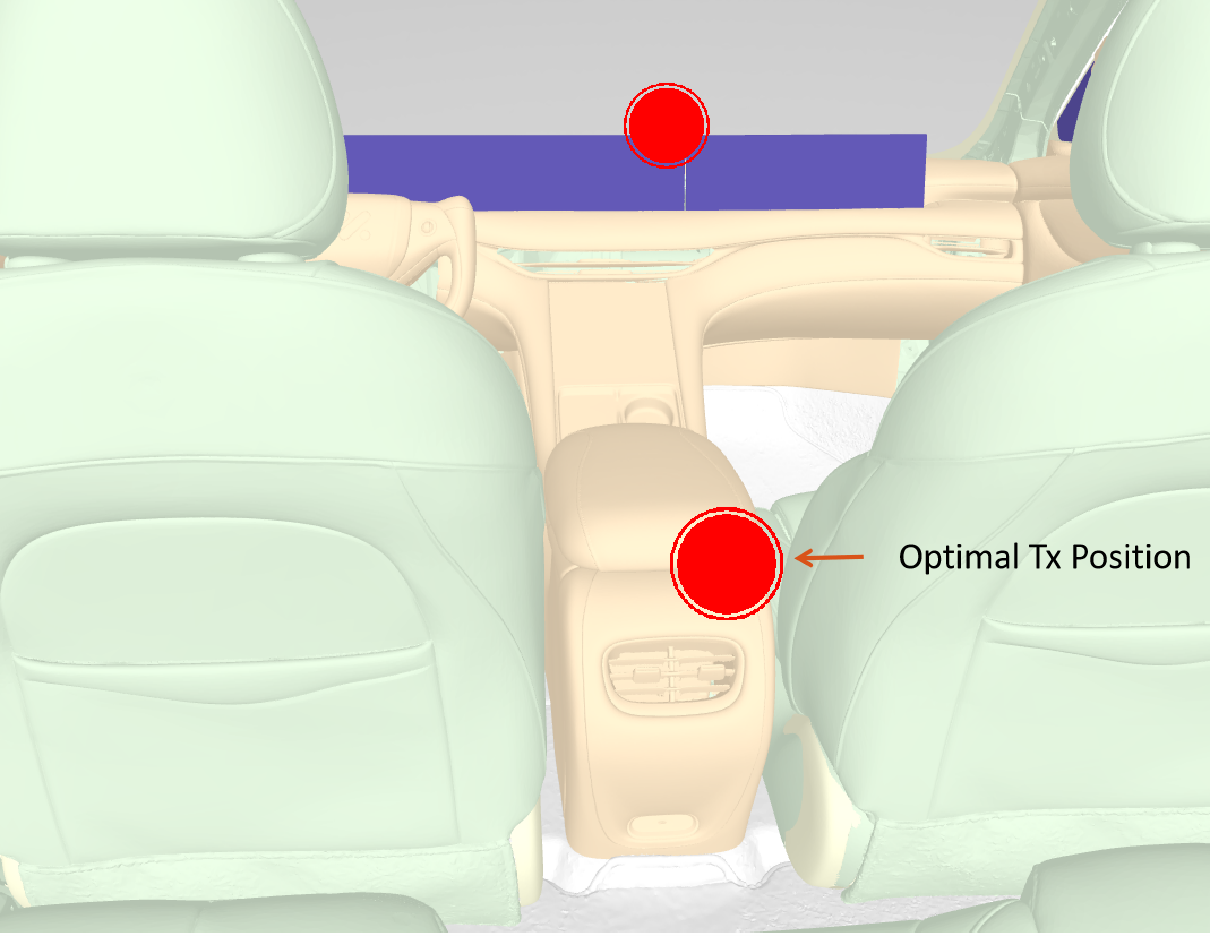}
}
\caption{Results after optimizing. (a) Coverage of different Tx. (b) Optimized position.}
\label{fig:WP16}
\end{figure}

A Powell–based optimizer is utilized to solve the maximum problem, considering the bound constraint. After optimization, an optimal Tx is found, as shown in Fig.~\ref{fig:WP16} (b). The coordinates of the point are $[1.933, 0.009, 0.763]$, obtained through an optimization process with Tx~4 as the initial reference point. To evaluate the coverage performance and practical feasibility of this location, its coverage probability versus threshold curve is plotted and shown in Fig.~\ref{fig:WP16} (a). As observed from the figure, compared to the previously identified optimal point Tx~4, the new point exhibits a higher coverage probability under the same threshold conditions. For example, at a threshold of $20$~dB, the coverage probability at Tx~4 drops below $30$\%, whereas the optimized point achieves a coverage probability exceeding $60$\%. Furthermore, the system rate can be raised up to $125.24$~Gbps. However, considering the spatial location of this point, it is situated at a certain distance from the nearest physical structure, which may pose challenges for real-world deployment. Therefore, this result is provided for reference only. An alternative position in the optimization procedure is found under the ceiling, balancing performance and feasibility, which has the potential to be the optimal antenna position in the deployment.

\subsection{Case B. Multiple Transmitters}

\begin{figure}[t]
\centering
\includesvg[width=0.9\columnwidth]{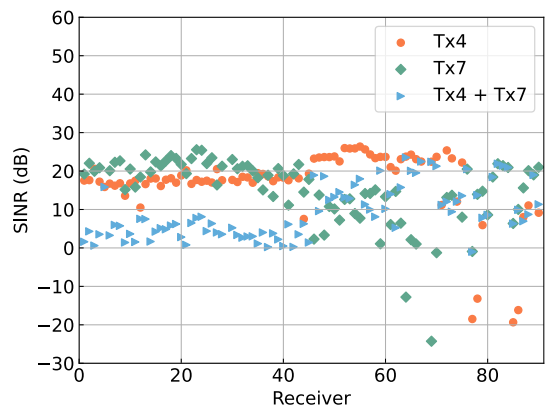}
\caption{SINR of different Tx (Multi-Tx Case).}
\label{fig:MaxRm}
\end{figure}

\begin{figure}[t]
\centering
4

\subfloat[]{
    \includesvg[width=0.9\columnwidth]{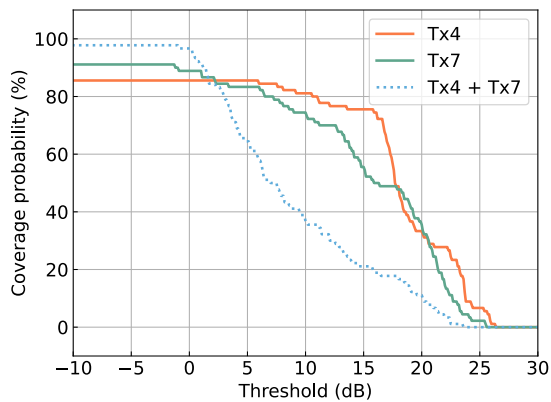}
}
\hfill
\subfloat[]{
    \includesvg[width=0.9\columnwidth]{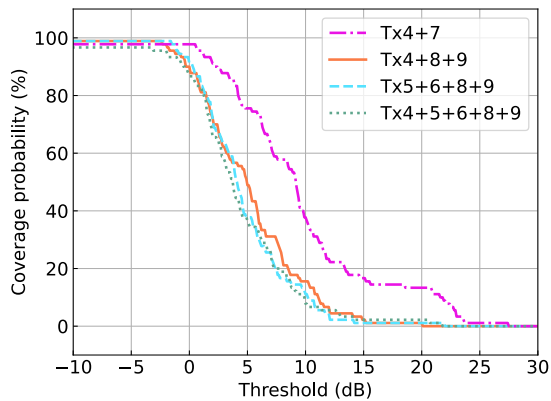}
}

\caption{Coverage probability of different Tx. (a) Single Tx vs Dual Txs (b) Different number of Txs.}
\label{fig:CPm}
\end{figure}

\begin{figure*}[t]
\centering
\subfloat[]{
    \includesvg[width=0.5\columnwidth]{figures/rm2_10.svg}
}
\subfloat[]{
    \includesvg[width=0.5\columnwidth]{figures/rm2_ass_10.svg}
}
\subfloat[]{
    \includesvg[width=0.5\columnwidth]{figures/rm3_10.svg}
}
\subfloat[]{
    \includesvg[width=0.5\columnwidth]{figures/rm3_ass_10.svg}
}
\hfill
\subfloat[]{
    \includesvg[width=0.5\columnwidth]{figures/rm4_10.svg}
}
\subfloat[]{
    \includesvg[width=0.5\columnwidth]{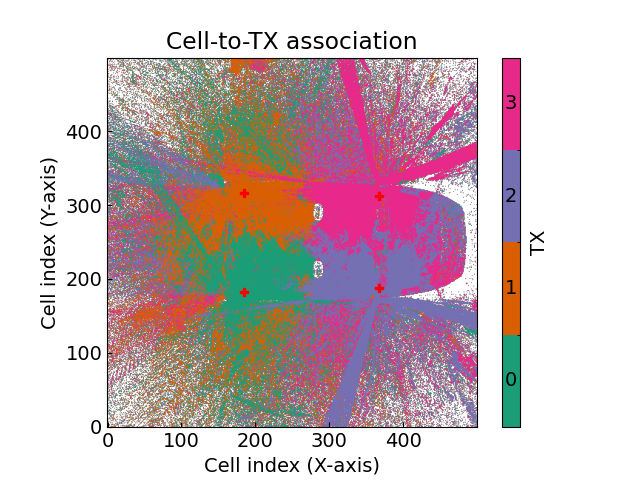}
}
\subfloat[]{
    \includesvg[width=0.5\columnwidth]{figures/rm5_10.svg}
}
\subfloat[]{
    \includesvg[width=0.5\columnwidth]{figures/rm5_ass_10.svg}
}

\caption{Coverage map of different Tx. (a)(c)(e)(g) SINR distribution of 2\,-\,5 Txs (b)(d)(f)(h) Assosiation of 2\,-\,5 Txs.}
\label{fig:CM}
\end{figure*}

\begin{table}[t]
\caption{Rate of different Tx (Multilple Txs Case).}
\label{tab:ratem}
\centering
\begin{tabular}{cc|cc}
\toprule
Tx & Rate\,/\,Gbps & Tx & Rate\,/\,Gbps\\
\midrule
~~~~4~~~~ & 108.42 & ~~~~4+8+9~~~~ & 43.91 \\
~~~~7~~~~ & 124.43 & ~~~~5+6+8+9~~~~ & 41.17 \\
~~~4+7~~~ & 69.47 & ~~~4+5+6+8+9~~~ & 40.44 \\
\bottomrule
\end{tabular}
\end{table}

When there is more than one Tx set in the scenarios, it is noteworthy whether the Txs would enlarge the coverage or interfere with each other more obviously. In this case, Tx~4 is placed beneath the ceiling at the center of the vehicle, while Tx~7 is positioned above the headrest in the middle of the rear seat. Coverage, interference, and rate analyses are carried out simultaneously.

\subsubsection{Interference from other Tx}
Considering multiple Txs in a car, where the Rx is connected to the nearest AP or the nearest LoS AP. Thus, other Txs act as interferers, causing a decrease in SINR. According to (\ref{equation: SINR}), SINR can be presented as
\begin{equation}
\begin{split}
    \mathrm{SINR} = \frac{P_tG_0K_ug(u)}{\sum_{i\in{\Phi_0/AP_0}}P_tG_iK_{x_i}g(x_{i}) + N_0B}.
\end{split}
\end{equation}

In the case of multiple Txs, the interference pattern changes. On one hand, as the Txs become denser, each UE can connect to a closer AP, resulting in higher received power. On the other hand, the number of other Txs acting as interference sources also increases, leading to stronger interference signals at each Rx. Therefore, simulation experiments are needed to evaluate how the increase in the number of APs affects the SINR, in order to determine the optimal number of Txs.

Fig.~\ref{fig:MaxRm} demonstrates the SINR of single Tx~4 (i.e., the optimal Tx in the front of the vehicle), Tx~7 (i.e., the compensatory Tx in the trunk), and the combination of Tx~4 and Tx~7. As shown in the figure, the SINR versus Rx position is plotted for the multi-Tx scenario. To better illustrate the impact of introducing multiple Txs on interference, the single-Tx cases are also included in the same figure for comparison. In the single-Tx scenario, the SINR of Tx7 is significantly higher than that of Tx4 at positions 1–45; however, the opposite trend is observed at other positions. In the Tx~4 + Tx~7 (i.e., multi-Tx) scenario, the SINR is generally lower than in the single Tx~4 case due to the interference introduced by the additional Tx. Only at certain Rx positions (e.g., positions 75–90) does the dual-Tx setup achieve higher SINR. But when compared to Tx~7, the joint of two Txs performs better. This indicates that, in most cases, the single-Tx setup demonstrates better performance, primarily because it avoids mutual interference between Txs. And it holds more significance to find out the optimal single Tx position, rather than adding more compensatory Txs.

\subsubsection{Coverage Probability}
When there is more than one Txs, coverage probability can still be calculated according to (\ref{equation: CP}), and the conditional coverage probability is
\begin{equation}
\begin{split}
    P_{c,L}(T) =\mathbb{P}(\frac{P_tG_0K_ug(u)}{\sum_{i\in{\Phi_0/AP_0}}P_tG_iK_{x_i}g(x_{i}) + N_0B}>T).
\end{split}
\end{equation}

Coverage of a multi-Tx case discloses the interference from other Tx and the propagation distance in the vehicle.
As is illustrated in Fig.~\ref{fig:CPm}, not only is the coverage-threshold curve of a single Tx plotted, but the coverage probability of two Txs combined is also acquired. Different number of Txs, including 2-5 Txs, is considered and plotted in Fig.~\ref{fig:CPm} (b).
Fig.~\ref{fig:CPm} (a) illustrates the variation of coverage probability with respect to the SINR threshold under both single- and dual-Tx scenarios. Taking Tx4 and Tx7 as examples, when the threshold is set to a relatively low value (below $0$ dB), the multi-Tx configuration can improve the coverage probability to almost 100\%. That is because some receiving points are in coverage blind spots when there is only one Tx, but can be covered by other Txs when multiple Txs are present. However, as the threshold increases above $2.5$ dB, the coverage probability under the multi-Tx setup decreases instead. This is due to the mutual interference caused by signals from different Txs. When the threshold is set above $25$ dB, the coverage probability of all the cases drops to almost 0, which is limited by the inherent propagation loss and transmission distance of THz waves, rather than by system-level network design.

Fig.~\ref{fig:CPm} (b) demonstrates the coverage ability as the number of Txs gradually increases in the vehicle system. As the number of Txs increases further, the improvement in coverage under low threshold (below $-2.5$ dB) becomes less significant, with coverage rates approaching nearly 100\%. However, under higher threshold values (above $2.5$ dB), the cumulative interference from additional Txs leads to a continuous decline in coverage probability at the same threshold. For example, with a threshold of 5~dB, two Txs can achieve a coverage probability of over 60\%, while scenarios with three or four Txs only reach less than 50\%, and the case with five Txs results in even lower coverage (around 20\%). This suggests that increasing the number of Txs does not significantly enhance the coverage area. Two Txs are sufficient to achieve full coverage in an in-vehicle scenario without causing excessive interference.

Fig.~\ref{fig:CM} represents the coverage map in multi-Tx case.The two Txs in this figure correspond to Tx~4 and Tx~7, both evaluated at a height of $z=1\,\mathrm{m}$. As shown in Fig.~\ref{fig:CM}(a), the overall SINR values are lower, with higher SINR regions primarily concentrated around each respective Tx. Fig.~\ref{fig:CM} (b) illustrates the connectivity at different locations within the vehicle. The front half of the vehicle connects to Tx~4 as the AP, while the rear half connects to Tx~7. This indicates that in a multi-Tx scenario, each Tx can effectively serve the nearby Rxs without the need to maintain high coverage quality over longer distances. The coverage map of more Tx cases demonstrates a similar result, with SINR getting lower and the association becoming more chaotic.

In summary, to enhance coverage and minimize the likelihood of communication dead zones, deploying two Txs as APs within the vehicle environment proves to be an effective strategy. Introducing more Txs may lead to increased mutual interference while offering limited improvement in coverage. Utilizing two Txs can effectively eliminate most coverage blind spots inside the vehicle while maintaining minimal interference, making it a potentially optimal configuration for multi-Tx scenarios.

\subsubsection{System Rate and Optimal Tx Analysis}
The average data rates under different scenarios are summarized in Table~\ref{tab:ratem}. In the multi-Tx case, the average data rate is calculated by averaging over all Txs. As shown in the table, the single Tx~$4$ configuration yields the highest data rate, which is attributed to its superior coverage probability. In contrast, the case where multiple Txs operate simultaneously results in relatively poor data rates and requires increased total transmit power due to the introduction of additional Txs. Therefore, the single-Tx configuration is preferable, and placing a Tx at the Tx~$4$ position (i.e., on the rooftop) provides the optimal communication performance.

\section{Conclusion}
\label{section:conclusion}
In this work, we presented a study on V2X THz wireless channel modeling based on DT-empowered RT simulations. A DT of the vehicle environment is constructed and integrated into an RT simulator, with the resulting data analyzed and visualized using PADPs. In both LoS and NLoS scenarios of with/without human and with/without window cases, the strong agreement between simulated and measured results validates the accuracy and practicality of the proposed simulation framework. In all cases, the measured results closely match the RT simulations, with errors remaining within an acceptable range. The main difference lies in the presence of a human inside the vehicle, which introduces attenuation to the OLoS path, leading to reduced received power. In contrast, rolling down the windows introduces minimal interference and has a negligible impact on signal propagation. Building on this validated RT model, we further investigated wireless system planning aspects, including SINR computation, coverage analysis, system rate, and optimal Tx placement. The results show that, in the single-Tx case, placing the Tx beneath the vehicle ceiling yields broader coverage and higher transmission rates, and in the multi-Tx case, the two-Tx case shows the best performance on coverage ability, providing valuable insights for the design and optimization of future in-vehicle THz communication systems.





\balance

\end{document}